\documentclass[final,5p,times,twocolumn]{elsarticle}

\usepackage{graphicx}
\usepackage{amssymb}
\usepackage{newtxtext, newtxmath, amsmath, fixme}
\usepackage{siunitx}
\DeclareSIUnit\sr{sr}
\usepackage{subcaption}
\usepackage{booktabs}
\usepackage{tabularx}
\usepackage{multirow}
\usepackage{enumitem}
\usepackage[version=4]{mhchem}
\usepackage{tikz}
\usetikzlibrary{arrows.meta,calc}
\usepackage[colorlinks=true, linkcolor=blue, urlcolor=blue, citecolor=blue]{hyperref}
\usepackage{cleveref}

\journal{Nuclear Instruments and Methods A}

\begin{document}
\begin{frontmatter}

    \title{Angle-Resolved Reflectance Measurements in the Vacuum Ultraviolet for Metallic Samples}

    \author[ific]{H.~Amar\corref{cor1}}
    \ead{Hamza.Amar@ific.uv.es}

    \author[ific]{J.~\'Alvarez}
    \author[ific]{X.~Arrue}
    \author[ific]{C.~Benitez}
    \author[ific]{A.~Cervera-Villanueva}
    \author[ific]{J.~Mart\'in-Albo}
    \author[ific]{L.~Molina-Bueno}
    \author[ific]{A.~Roche}
    \author[ific]{M.~Sorel}
    \author[ific]{J.~Soto-Oton\fnref{fn1}}
    \author[ific]{S.~Teruel-Pardo}
    \author[ific]{N.~Yahlali}

    \address[ific]{Instituto de F\'isica Corpuscular (IFIC), CSIC‐Universitat de Val\`encia, Spain}

    \cortext[cor1]{Corresponding author}

    \fntext[fn1]{Now at Nikhef National Institute of Subatomic Physics, Netherlands.}

    \begin{abstract}
        The light yield of noble-element scintillation detectors depends strongly on the reflective properties of detector materials in the vacuum ultraviolet (VUV), where optical measurements are scarce. For this reason, we built a custom rotary setup at IFIC, equipped with a monochromator and operated in a gaseous argon atmosphere, to perform angle-resolved reflectance measurements in the specular reflection plane. We describe the setup and its extensive characterization, together with the procedure used to measure the bidirectional reflectance distribution function (BRDF) of materials at various angles of incidence, for wavelengths from \qtyrange{128}{500}{\nm}, covering both the argon (\qty{128}{\nm}) and xenon (\qty{175}{\nm}) scintillation wavelengths. To illustrate the capabilities of the setup, we present results for rough aluminum and stainless-steel surfaces as reference materials. Fitting the measured angular profiles with the analytical GGX microfacet model allows us to extract the real and imaginary parts of the refractive index and the surface roughness parameter of each sample. These parameters are then used to extrapolate the in-plane measurements to the full hemisphere and from gaseous argon to liquid media, enabling the inference of the hemispherical reflectance in gaseous argon, liquid argon and liquid xenon.
    \end{abstract}
    

\end{frontmatter}

\tableofcontents
\newpage

\section{Introduction}
\label{sec:intro}

Dedicated measurements of the reflectance of detector materials are important for characterizing the response of particle detectors with optical readout, such as scintillation or Cherenkov detectors. The light yield, in detected photoelectrons per unit deposited energy, can generically be factorized into three terms, related to photon production, transport and detection, respectively:

\begin{equation}
    \label{eq:light_yield}
    \mathrm{LY} = \dfrac{N_\mathrm{ph}}{E_\mathrm{dep}} \, \dfrac{N_\mathrm{det}}{N_\mathrm{ph}} \, \mathrm{PDE}.
\end{equation}

The first term in \Cref{eq:light_yield}, $N_\mathrm{ph}/E_\mathrm{dep}$, is the number of optical photons produced in the detection medium per unit deposited energy. It depends on the particle type and on the detector configuration, for instance, on any electric field affecting electron–ion recombination and hence scintillation light production. The second term, $N_\mathrm{det}/N_\mathrm{ph}$, is the probability that an emitted photon reaches the active area of a photosensor or photon-detector module. It encompasses the detector geometrical acceptance and all aspects of photon transport, being the latter wavelength dependent. In particular, it is affected by scattering and absorption in the bulk of the detection medium, by shadowing from detector structures, and by the non-zero reflectance of detector materials. The latter aspect is the topic of this work. The third term, the photon detection efficiency (PDE), is the probability that a photon reaching the active area of a photosensor or photon-detector module produces a detectable photoelectron signal.

A vast literature exists on reflectance measurements in the visible range for the standard dielectric and metallic materials commonly used in particle detectors~\cite{polyanskiy2024}, including studies of different surface treatments (such as oxide layers) and aging effects. Measurements are much more scarce in the vacuum ultraviolet (VUV) region (\qtyrange{10}{200}{\nm}), which is where scintillation light is produced by noble elements. Particularly relevant for particle detection applications are the argon and xenon scintillation wavelengths, peaking at about \qty{127}{\nm} \cite{Heindl:2010zz} and \qty{175}{\nm} \cite{Fujii:2015qqq}, respectively. Moreover, even where measurements exist for relevant materials at relevant wavelengths, the angular distribution of the reflected light as a function of the incident direction is often not studied systematically. This matters for modeling photon transport and the resulting light yield, especially for reflectors that are neither perfectly diffuse nor perfectly specular, as is typically the case.

The experimental setup described and characterized in this work addresses this need by enabling angle-resolved reflectance measurements of detector materials from the VUV to the visible. Angular resolution is provided by a fully automated rotary system that scans both the incident and reflected light angles. The VUV regime, in turn, places strict requirements on the environment surrounding the light source, the sample and the photosensor. In VUV spectroscopy, vacuum is not just a convention but a functional requirement: VUV photons are energetic enough to induce electronic transitions and break molecular bonds, and are therefore strongly attenuated by atmospheric constituents, most notably oxygen (\ce{O2}) and water vapor (\ce{H2O}) \cite{Watanabe1953JCP, C3CP54696A}. Air must thus be removed from the optical path. For long-term operational robustness, a high-purity gas purge can be used instead of high vacuum. Molecular nitrogen is often chosen, but shows several absorption band systems in the \qtyrange{110}{145}{\nm} range, including the Lyman–Birge–Hopfield (LBH) bands \cite{lofthus1977spectrum}, and is therefore not optimal. We instead use a controlled gaseous argon (GAr) environment, rather than a standard vacuum chamber. This ensures VUV transparency above \qty{115}{\nm} while allowing flexibility to house complex instrumentation that could not operate in vacuum.

The paper is organized as follows. \Cref{sec:setup,sec:characterization} describe the experimental setup and its detailed characterization. \Cref{sec:measurement} presents the procedure used to measure the bidirectional reflectance distribution function (BRDF), and \Cref{sec:model} describes our BRDF modeling and the procedure to extract the hemispherical reflectance. Results for reference samples (in our case, rough aluminum and stainless-steel surfaces) are presented in \Cref{sec:results} and discussed in \Cref{sec:discussion} in the context of existing literature and particle detector applications.
\section{Experimental Setup}
\label{sec:setup}

\begin{figure}[!h]
    \centering
    \includegraphics[width=\columnwidth]{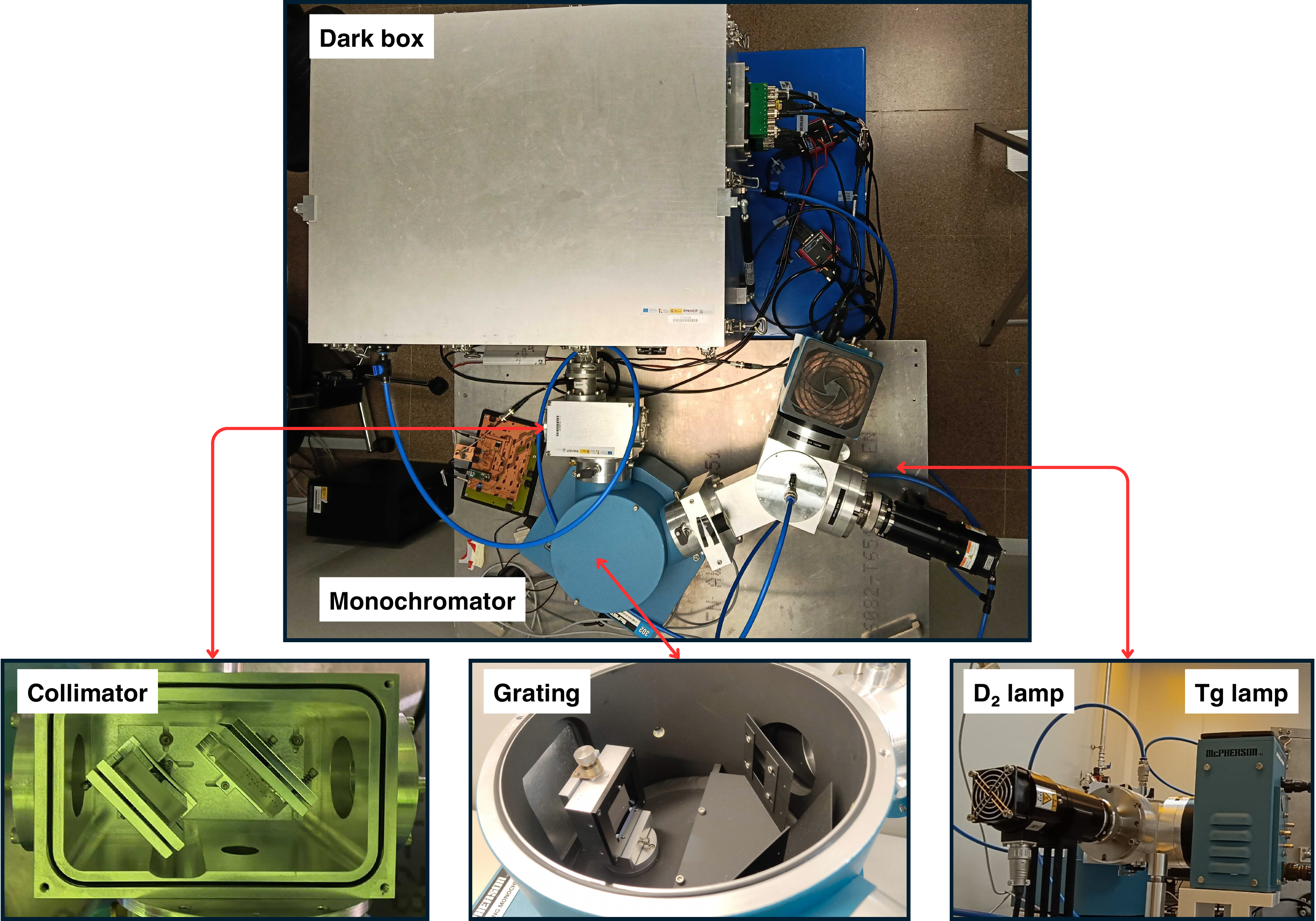}
    \caption{Setup for optical measurements in the VUV region. It comprises a deuterium and a tungsten lamp, a monochromator, and a VUV collimator.}
    \label{fig:monochromator}
\end{figure}

To precisely characterize the reflective properties of detector materials, a dedicated setup was developed inside a dark room. The light generation system couples both a deuterium ($\mathrm{D}_2$) lamp (Hamamatsu L15094) and a tungsten (Tg) lamp (McPherson 621) into a McPherson 302/234 monochromator (see \Cref{fig:monochromator}). Wavelength selection between \qtylist{115;550}{\nm} is achieved via a Seya-Namioka optical configuration, employing an aberration-corrected concave holographic diffraction grating with \qty{1200}{grooves/\mm}. To optimize reflectance in the VUV, the grating is coated with aluminum and enhanced with magnesium fluoride. The beam exiting the monochromator is collimated (McPherson VUV collimator with cylindrical mirrors and clear aperture of \qty{25}{mm}) and directed into an aluminum alloy dark box where the sample measurement takes place.

To enable light propagation in the VUV, the dark box holds a pure gaseous argon (GAr) atmosphere (Air Liquide ALPHAGAZ, \qty{99.999}{\percent} purity), which is continuously purged and maintained at a monitored overpressure of \qty{0.05}{bar}. This atmosphere also prevents the degradation of the grating, which is susceptible to the deposition of carbonaceous films originating from material outgassing under vacuum conditions~\cite{hollenshead2006modeling,dolgov2015characterization}.

\begin{figure}[!h]
    \centering
    \includegraphics[width=\columnwidth]{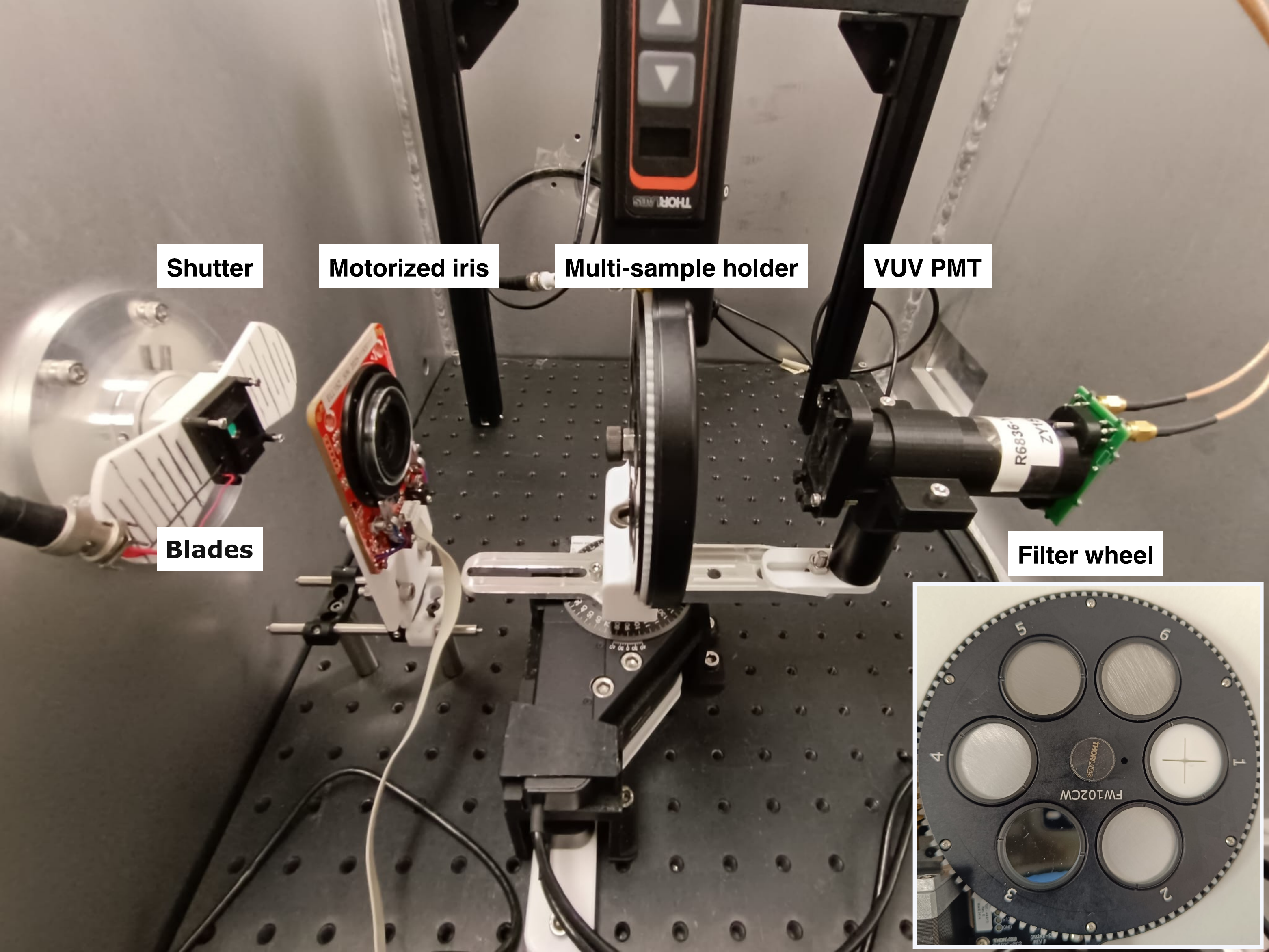}
    \caption{Setup for angle-resolved reflectance measurements. The 3D-printed assembly includes a motorized filter wheel for sample exchange and two rotation stages for a photosensor (top) and the multi-sample holder (bottom). A motorized iris reduces and collimates the beam spot while a shutter at the box entrance enables dark photocurrent measurements during scans.}
    \label{fig:reflectance_setup}
\end{figure}

Building upon the rotary setups described in \cite{Silva2010, kravitz2020}, angle-resolved reflectance measurements are conducted by means of a custom 3D-printed assembly positioned inside the dark box, as illustrated in \Cref{fig:reflectance_setup}. The manipulation relies on three automated motorized devices: a multi-sample holder (Thorlabs FW102C) controls sample exchange; a rotary stage (Thorlabs KPRM1E/M) regulates the position of a VUV-sensitive photomultiplier tube (PMT, Hamamatsu R6836-Y004) to set the detector angle; another rotary stage (Thorlabs KPRMTE/M) rotates the multi-sample holder to define the angle of incidence (AOI). The multi-sample holder is based on a filter wheel with six sample slots for 1-inch diameter samples. The two motorized stages share the same rotation axis; hence the angular alignment and the optical path are preserved, essential to avoid varying light attenuation from residual air contaminants at different positions during the purge.

The solid angle of the photosensor is restricted by a rectangular optical mask with the aperture placed directly in front of the PMT. The angular resolution is determined by the mask width and the distance with respect to the sample. The solid angle $\Omega$ subtended by a rectangular aperture of width $W$ and length $L$ at a distance $D$ from the sample center is evaluated as:
\begin{equation}
    \label{eq:mask_solid_angle}
    \Omega = 4 \arcsin \left( \frac{W \, L}{\sqrt{(W^2 + 4 D^2)(L^2 + 4 D^2)}} \right).
\end{equation}
With $W = \qty{2}{\mm}$, $L = \qty{10}{\mm}$, and $D = \qty{78.50}{\mm}$, \Cref{eq:mask_solid_angle} yields $\Omega \approx \qty{3.24e-3}{\sr}$, which corresponds to an angular resolution of \ang{1.46} in the specular reflection plane. More angular resolution can be achieved in detriment of the collected light intensity.

The PMT is operated in a gainless configuration and read with a picoammeter (Keithley 6487), while a \qty{+100}{\V} bias voltage is applied between the photocathode and the first dynode to ensure full photoelectron collection efficiency. A shutter, controlled by an Arduino microcontroller board, facilitates dark photocurrent measurements during data taking. Besides, a motorized iris (Thorlabs ELL15ZK) is positioned between the shutter and the rotary setup to reduce and further collimate the beam spot.

Finally, the alignment with the incident beam is verified at \qty{500}{\nm} by sequentially centering the beam spot on the iris aperture, a white dummy target in the multi-sample holder, and the PMT mask. Subsequently, the multi-sample holder angular offset for normal incidence is calibrated via a linear fit of the beam spot reflected off a mirror onto the white blades (\qty{0.5}{\cm} ticks) at the beam entrance (see \Cref{fig:reflectance_setup}). The setup alignment demonstrates effectiveness within an accuracy range of \numrange{1}{3} degrees.

To carry out the angle-resolved reflectance measurements under optimal conditions, a specific experimental configuration was established based on setup characterization studies. The collimation iris aperture is fixed at \qty{2.5}{\mm} to reduce the beam spot size on the sample while preserving enough light intensity. A calibration offset is applied to the multi-sample holder to align the sample normal with the incident beam. The PMT angular resolution (\ang{1.46}) defines the angular step when scanning the incident beam profile to avoid overlaps that would distort the total integrated photocurrent estimation. For the reflected beam profiles, a fine angular step equal to half of the angular resolution (\ang{0.73}) is used around the specular direction, whereas a coarser step of \ang{1.46} in the UV-VIS and \ang{2.92} in the VUV is adopted for the off-specular tails. The latter is set to minimize the gaseous argon purge time against potential purity drift.
\section{Setup Characterization} \label{sec:characterization}

\subsection{Monochromator Resolution}
\label{subsec:slits_characterization}

The monochromator bandwidth, in terms of the full width at half maximum (FWHM) was quantified via a Thorlabs CCS200/M spectrometer. The primary challenge in performing precise reflectance measurements, particularly for low reflective materials, is the light throughput: the lower the photocurrent, the more challenging the measurement. The maximum light throughput configuration, achieved with a \qty{3.0}{\mm} opening for both slits, yields a measured FWHM of \qty{12}{nm}. As long as no sharp absorption resonances are present, the variation of the sample reflectance within said bandwidth is small. Therefore, the reflectance measurements were recorded prioritizing light throughput.

\subsection{Lamp Spectra}
\label{subsec:lamp_spectra}

\begin{figure}[!h]
    \centering
    \includegraphics[width=\columnwidth]{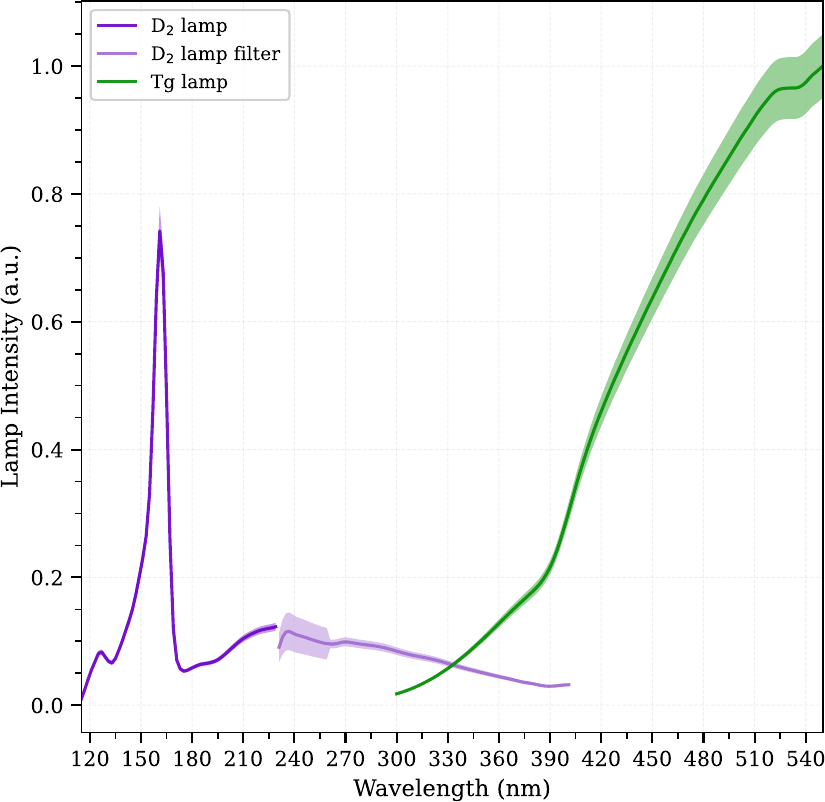}
    \caption{Deconvolved monochromator spectra for the $\mathrm{D}_2$ and Tg lamps after correcting for the photosensor quantum efficiency and filter transmittance, normalized to the maximum at \qty{550}{\nm}.}
    \label{fig:deconvolved_lamp_spectra}
\end{figure}

The deconvolved spectra of the monochromator, normalized to the maximum, under conditions of GAr saturation are depicted in \Cref{fig:deconvolved_lamp_spectra}, scanning the spectrum with a step of \qty{2}{\nm}. The $\mathrm{D}_2$ lamp spectrum (purple line) is divided into two wavelength intervals: \qtyrange{115}{230}{\nm} and \qtyrange{230}{400}{\nm}. The latter is measured with a \qty{225}{\nm} long-pass filter (Newport CGA-225), necessary to eliminate VUV second-order diffraction. The Tg lamp spectrum (green line), typically employed for optical measurements across most of the ultraviolet-visible (UV-VIS) range (from \qtyrange{300}{550}{\nm}), is included for comparison.

The calibrated PMT quantum efficiency (QE) and filter transmittance curves are employed for correction. A \qty{5}{\%} relative uncertainty is considered for both the QE and the plateau region of the long-pass filter. Conversely, a conservative \qty{25}{\%} uncertainty is assumed for the transition region, where a cut-on wavelength tolerance of \qty{5}{nm} is reported by the manufacturer. Within these tolerances and applying a Gaussian bandpass of \qty{12}{\nm}, the lamp spectra are deconvolved. In particular, the $\mathrm{D}_2$ lamp exhibits a consistent trend at \qty{230}{\nm}, free of VUV second-order diffraction contamination. The $\mathrm{D}_2$ and Tg lamp spectra enable reflectance measurements across extended spectral ranges of \qtyrange{115}{400}{\nm} and \qtyrange{300}{550}{\nm}, respectively. The overlapping region of \qtyrange{300}{400}{\nm} between the two lamps is utilized in subsequent analyses for internal cross-validation, as the reflectance of a sample at a given wavelength must remain consistent regardless of the light source.

\subsection{Gaseous Argon Atmosphere Stability}
\label{subsec:gar_purge_stability}

Prior to performing optical measurements in the VUV region, establishing a stable gaseous atmosphere is essential. Since a reflectance measurement involves a ratio between reflected and incident photocurrents taken at different times, temporal variations in the purity of the GAr atmosphere can bias the reflectance estimate. In particular, trace air contaminants, must be thoroughly purged.

\begin{figure}[!htbp]
    \centering
    \includegraphics[width=\columnwidth]{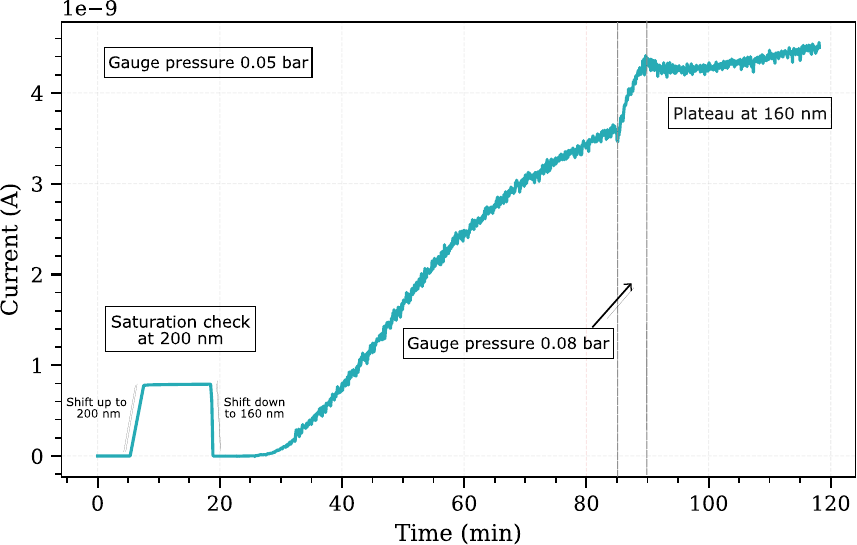}
    \caption{Monitoring of gaseous argon saturation within the dark box at \qty{160}{nm}. Transmission at \qty{200}{\nm} is used to check saturation at lower absorption levels.}
    \label{fig:gaseous_argon_saturation}
\end{figure}

To ensure precise measurements in the VUV, the photocurrent is continuously monitored with the VUV-sensitive PMT from the beginning of the GAr purge into the dark box. Monitoring is mainly performed at \qty{160}{\nm}, around the maximum of the spectrum (see \Cref{fig:deconvolved_lamp_spectra}), where absorption due to oxygen molecules is pronounced. As reported in \Cref{fig:gaseous_argon_saturation}, saturation is achieved in at most \qty{10}{\min} at \qty{200}{\nm}, where the absorption cross-section by air remnants is considerably lower. Upon \qty{30}{\min}, \qty{160}{\nm} light becomes visible, and a comparable photocurrent level is reached after \qty{40}{\min}.

To accelerate the purification of the atmosphere and thereby enhance light transmission, the gauge pressure is temporarily raised to \num{0.08} bar for \qty{5}{\min}, before restoring the nominal \num{0.05} bar overpressure for upcoming optical measurements. After \qty{2}{\hour}, the atmosphere is deemed suitable for recording reflectance measurements. Photocurrent variations are monitored both before and after the optical characterization of each sample via incident beam photocurrent measurements, and wavelengths most sensitive to purity variations are scanned last to maximize stability throughout the data acquisition.

\subsection{Validation with Specular Reference Standards}
\label{subsec:calibration_standards}

\begin{figure}[!h]
     \centering
     \includegraphics[width=\columnwidth]{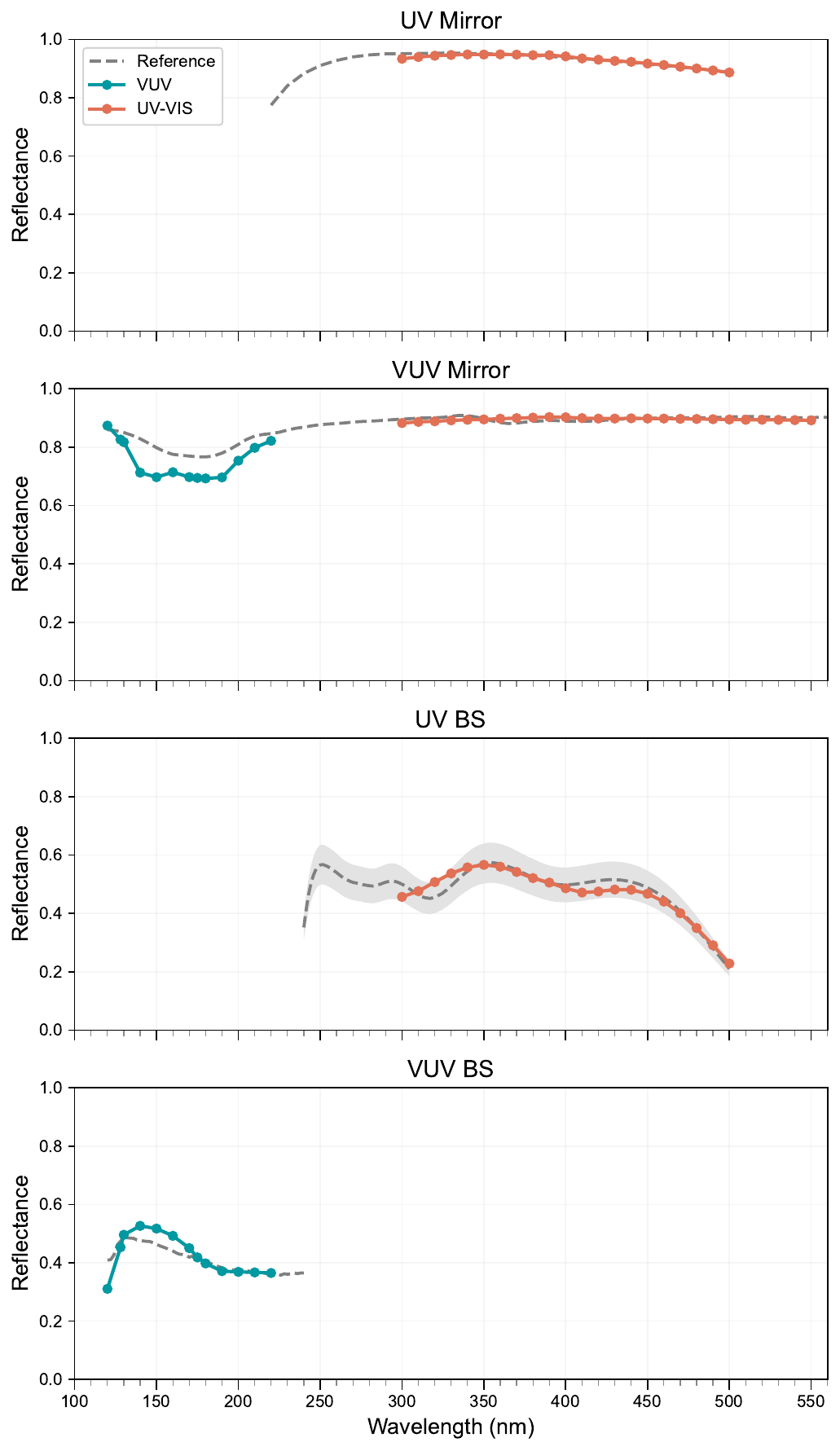}
     \caption{Calibration sample reflectances at \ang{45} AOI compared against manufacturer reference values (dashed lines). Only wavelengths without the presence of second-order diffraction were scanned. A \qty{12}{\%} reference uncertainty is indicated by shaded bands.}
     \label{fig:ref_reflectance}
\end{figure}

Finally, the measurement methodology with the rotary setup was validated across the VUV and UV-VIS ranges using four specular reference standards with known reflectance at \ang{45} AOI (\Cref{fig:ref_reflectance}): an aluminum UV mirror (Thorlabs PF10-03-F01), a magnesium fluoride-coated aluminum VUV mirror (Teledyne Acton Optics H1200-1D-MB), a UV beamsplitter (Thorlabs BSW20), and a VUV beamsplitter (Teledyne Acton Optics VUV-BS-45-1D). These calibration samples were measured without the photosensor mask so that the reflected beam impinged directly on the PMT window. Across all wavelengths dominated by first-order diffraction, the measured reflectances showed good agreement with the manufacturer reference curves. According to the manufacturer, a \qty{12}{\percent} tolerance is considered for the UV beamsplitter, while no tolerance is known for the rest.

\section{Reflectance Measurement Procedure}
\label{sec:measurement}

\subsection{The Bidirectional Reflectance Distribution Function}
\label{subsec:brdf_definition}

In optical radiometry, surface scattering is fundamentally described by the BRDF. It is defined as the ratio of the scattered radiance $dL_o$ propagating in an outgoing direction $\hat{\omega}_o$ (determined by the angles $\theta_o$ and $\phi_o$ in spherical coordinates) to the differential incident irradiance $dE_i$ illuminating the surface from the incoming direction $\hat{\omega}_i$~\cite{Nicodemus1992, Walter2007, Heitz2014}:
\begin{equation}
    \label{eq:exp_brdf}
    \text{BRDF}(\hat{\omega}_o, \hat{\omega}_i) = \frac{dL_o}{dE_i} = \frac{dL_o}{L_i \cos \theta_i d\omega_i},
\end{equation}
where $L_i$ is the incident radiance, $d\omega_i$ is the differential solid angle for the incoming direction. $\theta_i$ and $\theta_o$ denote the zenith angles of incidence and reflection relative to the macroscopic surface normal $\hat{\omega}_g$, while $\phi_i$ and $\phi_o$ are the azimuthal angles measured in the sample surface plane. For isotropic and non-magnetic materials, Stokes--Helmholtz reciprocity dictates that the BRDF is symmetric under exchange of incoming and outgoing trajectories: $\text{BRDF}(\hat{\omega}_o, \hat{\omega}_i) = \text{BRDF}(\hat{\omega}_i, \hat{\omega}_o)$~\cite{Nicodemus1992}.

Integrating the projected outgoing radiance across the upper hemisphere yields the directional-hemispherical reflectance $R(\hat{\omega}_i)$~\cite{Nicodemus1992}:
\begin{equation}
    \label{eq:hemis_reflectance}
    R(\hat{\omega}_i) = \int\limits_{\theta_o=0}^{\frac{\pi}{2}}\int\limits_{\phi_o= 0}^{2\pi} \text{BRDF}(\hat{\omega}_o, \hat{\omega}_i) \, \cos \theta_o \, \sin \theta_o \, d\theta_o \, d\phi_o,
\end{equation}
For an isotropic reflective surface, the hemispherical reflectance at a certain wavelength depends exclusively on the angle of incidence $\theta_i$.

Experimentally, the BRDF is determined from the measured photodetector current in the plane of specular reflection as:
\begin{equation}
    \label{eq:brdf_experimental_formula}
    \text{BRDF}(\theta_o, \theta_i) = \frac{I_{\text{sample}}(\theta_o, \theta_i) - I_{\text{dark}}}{\left(I_{\text{i}} - I_{\text{dark}}\right) \, \cos\theta_o \, \Omega },
\end{equation}
where $I_{\text{sample}}(\theta_o, \theta_i)$ is the photocurrent measured along the outgoing direction, $I_{\text{i}}$ is the reference incident beam photocurrent used for normalization, $I_{\text{dark}}$ is the dark photocurrent, and $\Omega$ is the solid angle of the photosensor mask defined in \Cref{eq:mask_solid_angle}.

\subsection{BRDF Data Preparation}
\label{subsec:data_preparation}

To compute integrated reflectances from the angle-resolved measurements, raw photocurrents are first evaluated to reconstruct the experimental BRDF according to \Cref{eq:brdf_experimental_formula}, which is subsequently analyzed via two complementary paths: a model-independent linear-spline integration over the measurement plane to extract the in-plane reflectance $R_{\text{in-plane}}$ and the specular peak width (\Cref{subsec:in_plane_reflectance}), and a non-linear fit to the GGX microfacet model (\Cref{sec:model}) to determine the hemispherical reflectance, compared against literature benchmarks~\cite{polyanskiy2024}, and extrapolate the reflectance to liquid argon and liquid xenon.

Before evaluating \Cref{eq:brdf_experimental_formula}, the dark photocurrent recorded at each angular position is subtracted from both the reference incident beam and sample reflection scans. Because the dark box is continuously purged with gaseous argon, VUV transparency gradually improves during data acquisition. To prevent this drift from biasing the normalization, the reference incident photocurrent $I_{\text{i}}$ is measured before and after each sample scan and interpolated in time using a cubic spline. Accounting for this time-dependent incident intensity shifts the reconstructed BRDF downward, preventing a systematic overestimation of the VUV reflectance.

To reject noise-dominated points at large off-specular scattering angles, only positive net currents satisfying a threshold of $\text{SNR} \ge 6$ relative to their statistical uncertainty are analyzed. This is particularly important in the VUV, where the peak signal-to-noise ratio (SNR) is much lower due to lower light intensity and sample reflectance (e.g., it is approximately 40 times lower at \qty{175}{\nm} than at \qty{500}{\nm}).

Finally, systematic uncertainties in the BRDF, which stem from sample alignment and tilt, detector solid angle estimation, motorized stage repeatability, and beam intensity fluctuations, are quantified through 16 scans (718 measurement points) recorded under identical conditions. The relative spread, defined as the standard deviation divided by the mean BRDF at each detector angle $\theta_r$\footnote{$\theta_r$ denotes the raw angular position of the photosensor, while $\theta_o$ is the actual reflection after systematic corrections, as highlighted later in \Cref{sec:model}.}, is driven by motorized repositioning and alignment near the specular peak, while it is dominated by dark photocurrent fluctuations in the low-signal tails. Parameterizing this spread as a function of the signal-to-dark-photocurrent ratio ($\text{SDCR} = |I_{\text{net}}| / I_{\text{dc}}$) provides a wavelength-inclusive power-law fit that defines the relative uncertainty considered for the BRDF data:
\begin{equation}
    \label{eq:BRDF_systematics}
    \frac{\sigma_{\scalebox{0.55}{\text{BRDF}}}}{\text{BRDF}} = 0.119 \: \text{SDCR}^{-0.40},
\end{equation}
with an upper ceiling of \qty{30}{\percent} imposed to avoid unphysical divergence in low-signal regimes.

\subsection{Extraction of In-Plane Reflectance}
\label{subsec:in_plane_reflectance}

To evaluate the fraction of reflected light collected within the specular reflection plane without relying on a surface microfacet model, a model-independent in-plane reflectance $R_{\text{in-plane}}$ is estimated by integrating the discrete BRDF measurements, restricting the integration range in \Cref{eq:hemis_reflectance} to the experimental scattering angles.

Because the photodetector slit has a finite length $L$ perpendicular to the specular reflection plane, the measurements capture light scattered over a narrow, zenith-angle-dependent out-of-plane azimuthal range $[-\Delta \phi_r(\theta_r), \Delta \phi_r(\theta_r)]$, where:
\begin{equation}
    \label{eq:app_delta_phi}
    \Delta \phi_r(\theta_r) = \arctan\left(\frac{L}{2 D \sin\theta_r}\right),
\end{equation}
with $D$ being the sample-to-detector distance. In this case, the in-plane reflectance is written as:
\begin{equation}
    \label{eq:app_inplane_single_integral}
    R_{\text{in-plane}} = \int_{\theta_{r, \min}}^{\theta_{r, \max}} \text{BRDF}(\theta_r) \cos\theta_r \sin\theta_r \, 2 \Delta \phi_r(\theta_r) \, d\theta_r.
\end{equation}

To evaluate \Cref{eq:app_inplane_single_integral} from discrete measurements, a continuous profile defined by linear splines is used to compute the nominal integrated reflectance since they provide a robust, monotonic, and physically consistent representation. In contrast, while cubic splines produce continuously differentiable curves, they introduce unphysical oscillations near sharp gradients or in regions with scarce data after SNR filtering. A linear spline interpolation is also employed to characterize the specular peak width. Because the spline interpolation and subsequent numerical operations do not present simple analytical representations, the propagation of uncertainties is performed numerically using central finite differences.
\section{BRDF Modeling and Extraction of Hemispherical Reflectance}
\label{sec:model}

To describe light scattering from rough surfaces, microfacet theory conceives a flat macroscopic surface as an ensemble of microscopic specular facets whose orientations follow a statistical distribution of surface normals~\cite{Walter2007, Heitz2014, CookTorrance1982}. Under this geometric-optics approximation, each microfacet reflects incident light strictly according to Snell's law. Consequently, reflection from an incident direction $\hat{\omega}_i$ to an outgoing direction $\hat{\omega}_o$ occurs only on microfacets whose local normal $\hat{\omega}_m$ aligns with the half-vector $\hat{\omega}_h$, defined as the angular bisector of the incoming and reflected directions:
\begin{equation}
    \label{eq:half_vector}
    \hat{\omega}_h = \frac{\hat{\omega}_i + \hat{\omega}_o}{|\hat{\omega}_i + \hat{\omega}_o|}.
\end{equation}

The specular BRDF implemented to analyze the angular reflectance scans is formulated as:
\begin{equation}
    \label{eq:BRDF_generic}
    \text{BRDF}(\hat{\omega}_o, \hat{\omega}_i; n, \kappa, \gamma) = \frac{F(\theta'; n, \kappa) \, G_2(\theta_i, \theta_o; \gamma) \, D(\theta_h; \gamma)}{4 \cos \theta_i \cos \theta_o}.
\end{equation}
\Cref{eq:BRDF_generic} incorporates the real refractive index $n$ and extinction coefficient $\kappa$ of the metallic sample, which determine the complex refractive index $\tilde{n} = n + i\kappa$ in the Fresnel reflection coefficient $F(\theta'; n, \kappa)$ (\Cref{subsec:fresnel_reflections}). In addition, the dimensionless surface roughness parameter $\gamma$ parametrizes both the distribution of microfacet normals $D(\theta_m; \gamma)$ and the geometric shadowing-masking function $G_2(\theta_i, \theta_o; \gamma)$.

\begin{figure}[!h]
    \centering
    \includegraphics[trim={21.4cm 0 0 0}, clip, width=\columnwidth]{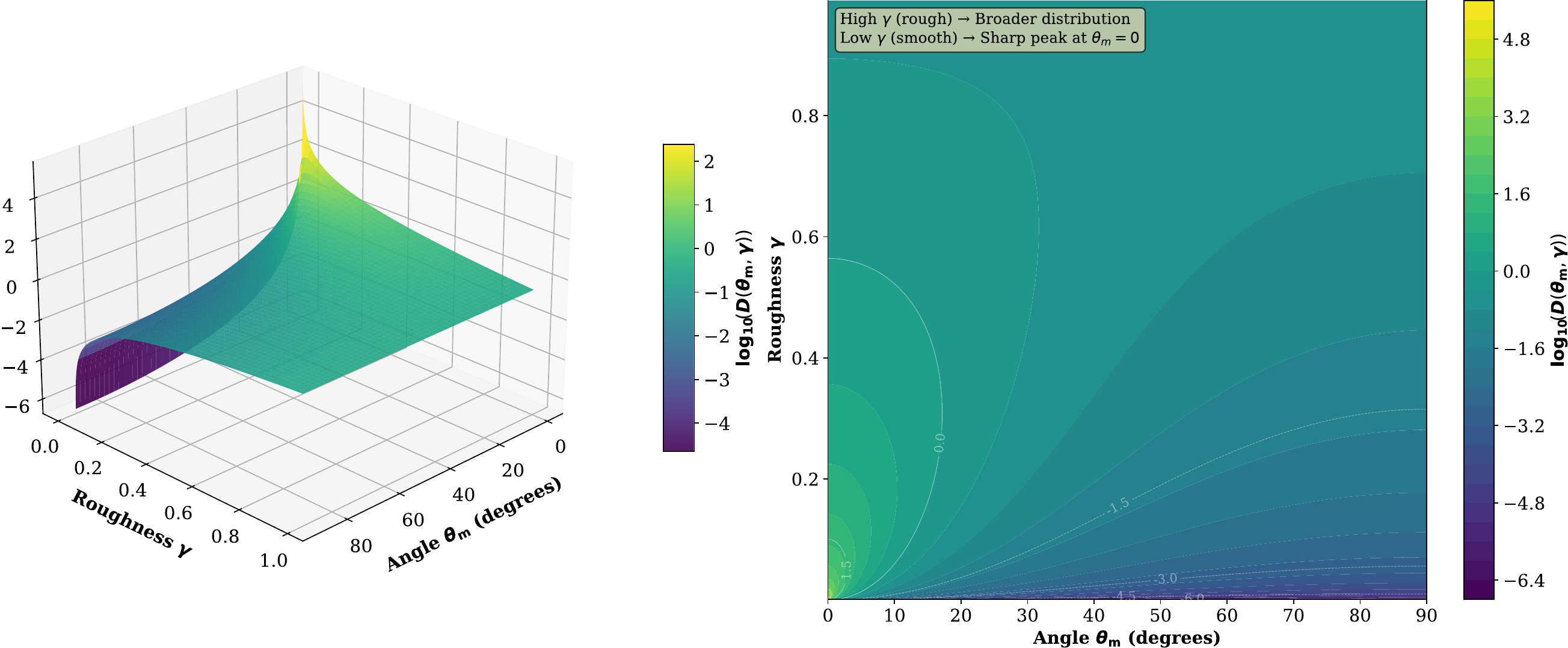}
    \caption{GGX normal distribution function $D$ as a function of the roughness parameter $\gamma$ and microfacet angle $\theta_m$.}
    \label{fig:GGX_dist}
\end{figure}

The distribution of microfacet normals employed in this study is the GGX distribution~\cite{Walter2007}, equivalent to the Trowbridge--Reitz distribution~\cite{Trowbridge:75}:
\begin{equation}
    \label{eq:GGX_microdistribution}
    D(\theta_m; \gamma) = \frac{\gamma^2 \chi^+(\cos \theta_m)}{\pi \cos^4 \theta_m \left( \gamma^2 + \tan^2 \theta_m \right)^2},
\end{equation}
where $\chi^+$ is the Heaviside step function and $\theta_m$ is the angle between the microfacet normal $\hat{\omega}_m$ and the macroscopic normal $\hat{\omega}_g$:
\begin{equation}
    \label{eq:microfacet-angle}
    \cos \theta_m = \frac{\cos \theta_i + \cos \theta_o}{\sqrt{2 \left( 1 + \cos \theta_i \cos \theta_o + \sin \theta_i \sin \theta_o \cos(\phi_o - \phi_i) \right)}}.
\end{equation}
As illustrated in \Cref{fig:GGX_dist}, the GGX distribution is parameterized by the dimensionless roughness variable $\gamma$. It features a prominent specular peak ($\theta_m \simeq 0$) accompanied by heavy tails, reproducing the scattering lobes observed experimentally for rough metallic surfaces. Because the distribution is isotropic, it depends only on the zenith angle $\theta_m$ and is independent of the azimuthal orientation $\phi_m$.

\begin{figure}[!h]
    \centering
    \begin{subfigure}[b]{0.48\textwidth}
        \centering
        \begin{tikzpicture}[
                >=Stealth,
                scale=0.82,
                rayIn/.style={thick, color=teal!70!black, ->},
                rayOut/.style={thick, color=teal!90!black, ->},
                blockedRay/.style={thick, dashed, color=red!80!black},
                blockedRayOut/.style={thick, dashed, color=teal!40},
                surface/.style={thick, draw=black!75, fill=gray!20},
                normalVec/.style={thick, color=black!70, ->}
            ]
            \fill[surface]
            (0,0) -- (0,1.0)
            to[out=30, in=170] (1.8, 2.2)
            to[out=-10, in=190] (3.8, 0.9)
            to[out=10, in=170] (5.2, 2.1)
            to[out=-10, in=180] (7.0, 1.2)
            -- (7.0,0) -- cycle;
            \draw[thick, draw=black!80]
            (0,1.0)
            to[out=30, in=170] (1.8, 2.2)
            to[out=-10, in=190] (3.8, 0.9)
            to[out=10, in=170] (5.2, 2.1)
            to[out=-10, in=180] (7.0, 1.2);

            \coordinate (M1) at (3.8, 0.9);
            \draw[ultra thick, color=red!80!black] (3.5, 0.85) -- (4.1, 0.95) coordinate[midway] (facet1);
            \draw[normalVec] (facet1) -- ++(-0.13, 0.78) node[above, font=\footnotesize] {$\hat{\omega}_h$};

            \coordinate (Peak1) at (1.8, 2.2);

            \draw[rayIn, <-] (0.5, 3.0) -- (Peak1) node[midway, above, font=\footnotesize] {$\hat{\omega}_i$};
            \draw[blockedRay] (Peak1) -- (facet1);
        \end{tikzpicture}
        \caption{Shadowing}
        \label{subfig:shadowing}
    \end{subfigure}
    \hfill
    \begin{subfigure}[b]{0.48\textwidth}
        \centering
        \begin{tikzpicture}[
                >=Stealth,
                scale=0.82,
                rayIn/.style={thick, color=teal!70!black, ->},
                rayOut/.style={thick, color=teal!90!black, ->},
                blockedRay/.style={thick, dashed, color=red!80!black},
                blockedRayOut/.style={thick, dashed, color=teal!40},
                surface/.style={thick, draw=black!75, fill=gray!20},
                normalVec/.style={thick, color=black!70, ->}
            ]
            \fill[surface]
            (0,0) -- (0,1.2)
            to[out=20, in=190] (1.8, 1.5)
            to[out=10, in=170] (3.2, 0.9)
            to[out=-10, in=190] (5.2, 2.3)
            to[out=10, in=180] (7.0, 1.1)
            -- (7.0,0) -- cycle;
            \draw[thick, draw=black!80]
            (0,1.2)
            to[out=20, in=190] (1.8, 1.5)
            to[out=10, in=170] (3.2, 0.9)
            to[out=-10, in=190] (5.2, 2.3)
            to[out=10, in=180] (7.0, 1.1);

            \coordinate (M2) at (3.2, 0.9);
            \draw[ultra thick, color=red!80!black] (2.9, 0.95) -- (3.5, 0.85) coordinate[midway] (facet2);
            \draw[normalVec] (facet2) -- ++(0.13, 0.78) node[above, font=\footnotesize] {$\hat{\omega}_h$};

            \draw[rayIn, <-] (1.92, 2.52) -- (facet2) node[midway, left, font=\footnotesize] {$\hat{\omega}_i$};

            \coordinate (Peak2) at (5.2, 2.3);

            \draw[rayOut] (facet2) -- (4.4, 1.74);
            \draw[blockedRay] (4.4, 1.74) -- (Peak2);
            \draw[blockedRayOut, ->] (Peak2) -- ++(1.1, 0.77) node[right, font=\footnotesize] {$\hat{\omega}_o$};
        \end{tikzpicture}
        \caption{Masking}
        \label{subfig:masking}
    \end{subfigure}
    \caption{Schematic representation of geometric occlusions on a rough surface profile: (a) shadowing, where incident rays are blocked by neighboring microfacets; (b) masking, where reflected rays toward the outgoing direction are obstructed by adjacent microfacets.}
    \label{fig:shadowing_masking_schematic}
\end{figure}
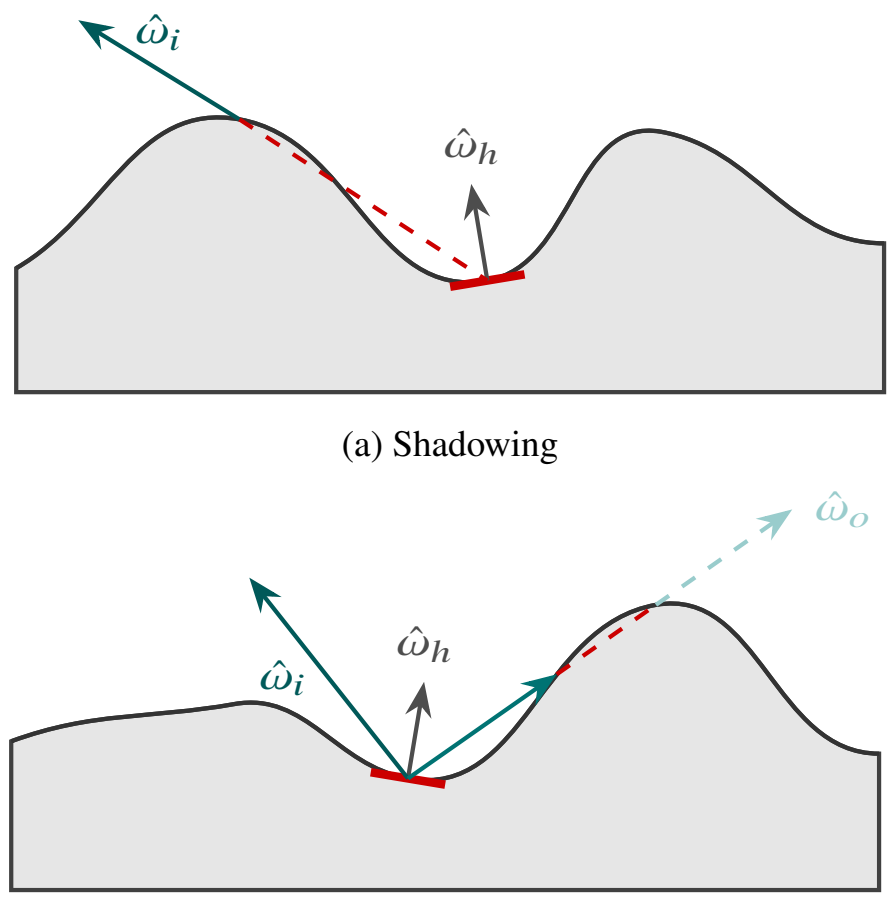

Geometrical occlusions among neighboring microfacets are modeled by the bidirectional masking-shadowing function $G_2(\theta_i, \theta_o; \gamma)$. On a rough surface, elevated microfacets obstruct incoming rays from reaching other facets (shadowing) or block reflected rays along the trajectory toward the detector (masking), as schematized in \Cref{fig:shadowing_masking_schematic}. In the GGX model, the Smith height-correlated masking-shadowing function is adopted~\cite{Walter2007, Heitz2014}:
\begin{equation}
    \label{eq:masking-shadowing-g2}
    G_2(\theta_i, \theta_o; \gamma) = \frac{2 \, \chi^+(\hat{\omega}_o \cdot \hat{\omega}_h) \, \chi^+(\hat{\omega}_i \cdot \hat{\omega}_h)}{\sqrt{1 + \gamma^2 \tan^2\theta_i} + \sqrt{1 + \gamma^2 \tan^2\theta_o}}.
\end{equation}
This formulation accounts for correlation between shadowing and masking without assuming statistical independence between the incident and viewing directions and precludes multiple scattering.

\subsection{Fresnel Reflections at Conductive Boundaries}
\label{subsec:fresnel_reflections}

At each microfacet mirror, the reflection amplitude is ruled by electromagnetic boundary conditions described by Maxwell's equations. For an unpolarized incident beam passing through a non-absorptive gaseous argon atmosphere ($n_0 \approx 1$) and impinging on a conductive metal characterized by complex refractive index $\tilde{n} = n + i\kappa$, the Fresnel reflection coefficient is evaluated as the arithmetic mean of the transverse electric ($s$-polarized) and transverse magnetic ($p$-polarized) components~\cite{born1999principles, FresnelCoefficients}:
\begin{equation}
    \label{eq:fresnel_unpolarized}
    F(\theta'; n, \kappa) = \frac{1}{2} \left[ R_s(\theta'; n, \kappa) + R_p(\theta'; n, \kappa) \right].
\end{equation}
$\theta'$ represents the angle of incidence between the incident vector $\hat{\omega}_i$ and the microfacet normal $\hat{\omega}_h$, extracted from:
\begin{equation}
    \label{eq:microfacet-refl-angle}
    \cos 2\theta' = \cos \theta_i \cos \theta_o + \sin \theta_i \sin \theta_o \cos(\phi_o - \phi_i).
\end{equation}

For metallic conductors, the polarization-dependent reflectances $R_s$ and $R_p$ are given by~\cite{born1999principles, FresnelCoefficients}:
\begin{align}
    \label{eq:fresnel_rs}
    R_s(\theta^\prime) & = \frac{a^2 + b^2 - 2a\cos\theta^\prime + \cos^2\theta^\prime}{a^2 + b^2 + 2a\cos\theta^\prime + \cos^2\theta^\prime},                                                                                            \\[0.5em]
    \label{eq:fresnel_rp}
    R_p(\theta^\prime) & = R_s(\theta^\prime) \frac{a^2 + b^2 - 2a\sin\theta^\prime\tan\theta^\prime + \sin^2\theta^\prime\tan^2\theta^\prime}{a^2 + b^2 + 2a\sin\theta^\prime\tan\theta^\prime + \sin^2\theta^\prime\tan^2\theta^\prime},
\end{align}
where the auxiliary variables $a$ and $b$ are defined as:
\begin{align}
    a^2 & = \frac{1}{2n^2_0} \Biggl[ \sqrt{\left( n^2 - \kappa^2 - n^2_0 \sin^2\theta^\prime \right)^2 + 4 n^2 \kappa^2} \nonumber \\
        & \qquad + \left( n^2 - \kappa^2 - n^2_0 \sin^2\theta^\prime \right) \Biggr], \label{eq:fresnel_a} \\[0.5em]
    b^2 & = \frac{1}{2n^2_0} \Biggl[ \sqrt{\left( n^2 - \kappa^2 - n^2_0 \sin^2\theta^\prime \right)^2 + 4 n^2 \kappa^2} \nonumber \\
        & \qquad - \left( n^2 - \kappa^2 - n^2_0 \sin^2\theta^\prime \right) \Biggr], \label{eq:fresnel_b}
\end{align}
with $n_0$ being the refractive index of the immersion medium.

The Fresnel coefficient in~\Cref{eq:fresnel_unpolarized} modulates the overall amplitude of the specular lobe and governs its amplification as the angle of incidence increases toward grazing angles. \Cref{eq:BRDF_generic} does not include any diffuse term because the extinction coefficient prevents internal scattering.

To fit the GGX model to the experimental angular scans, the outgoing zenith angle is parameterized as $\theta_o = \theta_r - \theta_s$, where $\theta_s$ is a free offset common to all angular scans for a given sample and wavelength, accounting for minor alignment shifts between the beam axis and the rotary setup. Because the photodetector subtends a finite solid angle, the theoretical BRDF is numerically convolved via 2D trapezoidal integration over the rectangular mask aperture. The model parameters are then extracted by $\chi^2$ minimization. Once the best-fit parameters are determined, the hemispherical reflectance is computed numerically according to \Cref{eq:hemis_reflectance}.
\section{Results}
\label{sec:results}

To evaluate the reflective performance of materials relevant to noble-element particle detectors, two metallic samples of 1-inch diameter were characterized with the rotary setup in \Cref{fig:reflectance_setup}: a bare aluminum sample (AW-1050~(A)) with average roughness ($R_a$ parameter) of \qty{0.362}{\micro\meter} and a stainless steel sample (SS 304) with average roughness of \qty{0.374}{\micro\meter}.

\subsection{BRDF Measurements and In-Plane Reflectance}
\label{subsec:brdf_measurements}

\begin{figure*}[!h]
    \centering
    \begin{minipage}{0.85\textwidth}
        \begin{subfigure}[b]{0.49\linewidth}
            \centering
            \includegraphics[width=\linewidth]{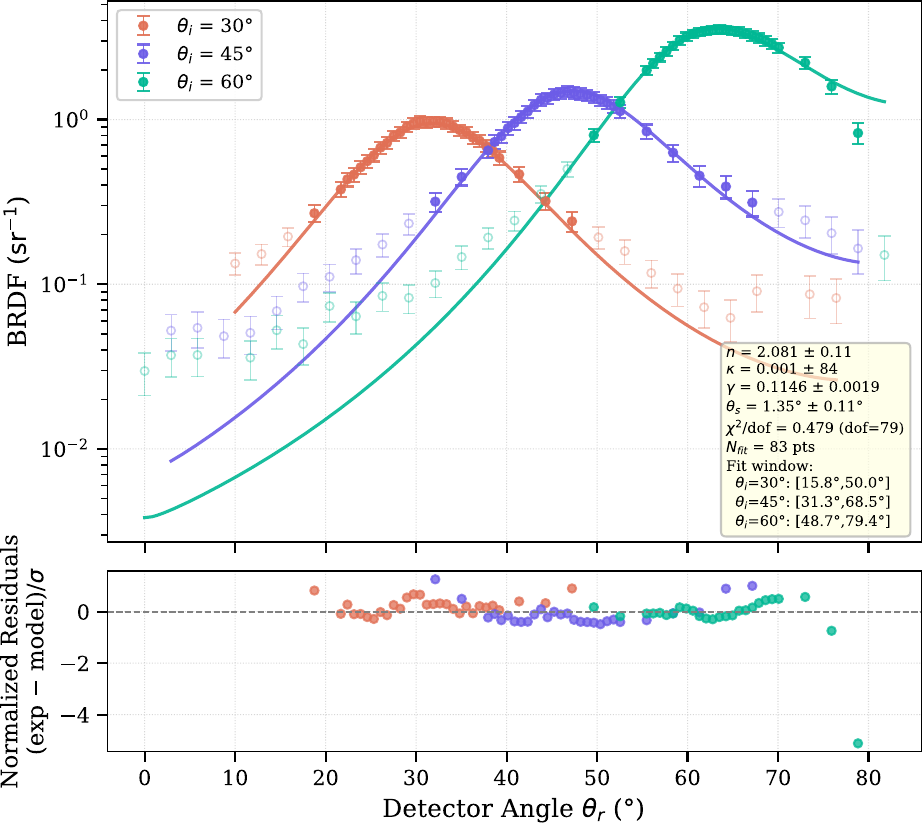}
            \caption{Al \qty{0.362}{\micro\meter}, \qty{175}{\nm}}
            \label{fig:brdf_fit_al_175}
        \end{subfigure}
        \hfill
        \begin{subfigure}[b]{0.49\linewidth}
            \centering
            \includegraphics[width=\linewidth]{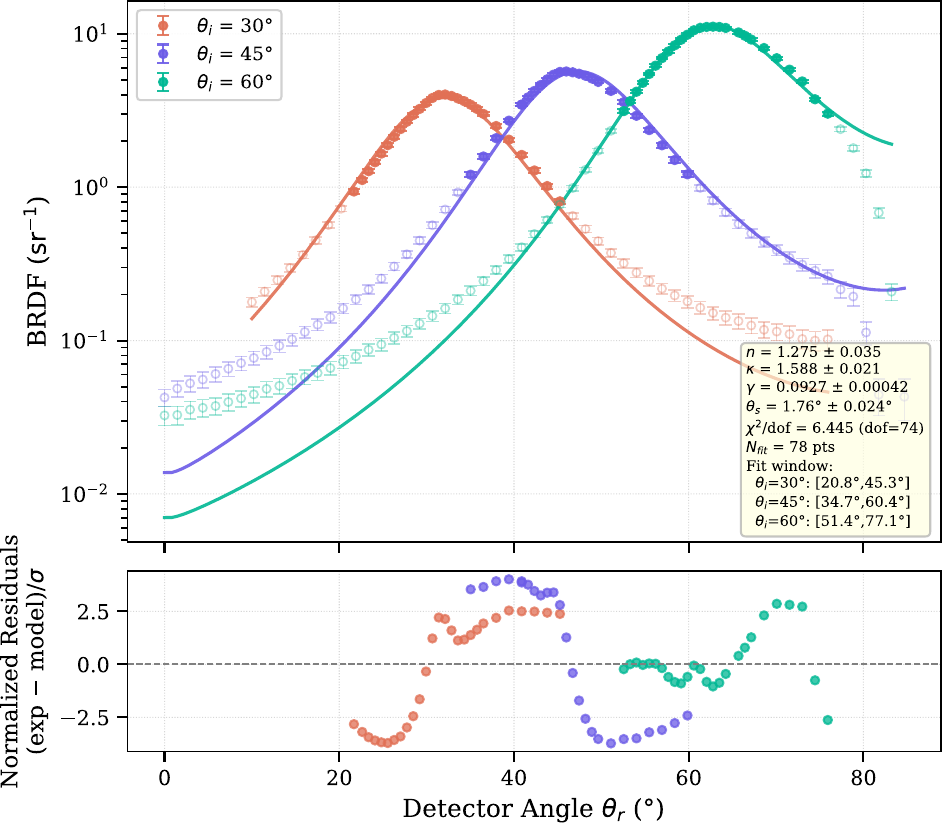}
            \caption{Al \qty{0.362}{\micro\meter}, \qty{500}{\nm}}
            \label{fig:brdf_fit_al_500}
        \end{subfigure}
    
        \vspace{0.5cm}
    
        \begin{subfigure}[b]{0.49\linewidth}
            \centering
            \includegraphics[width=\linewidth]{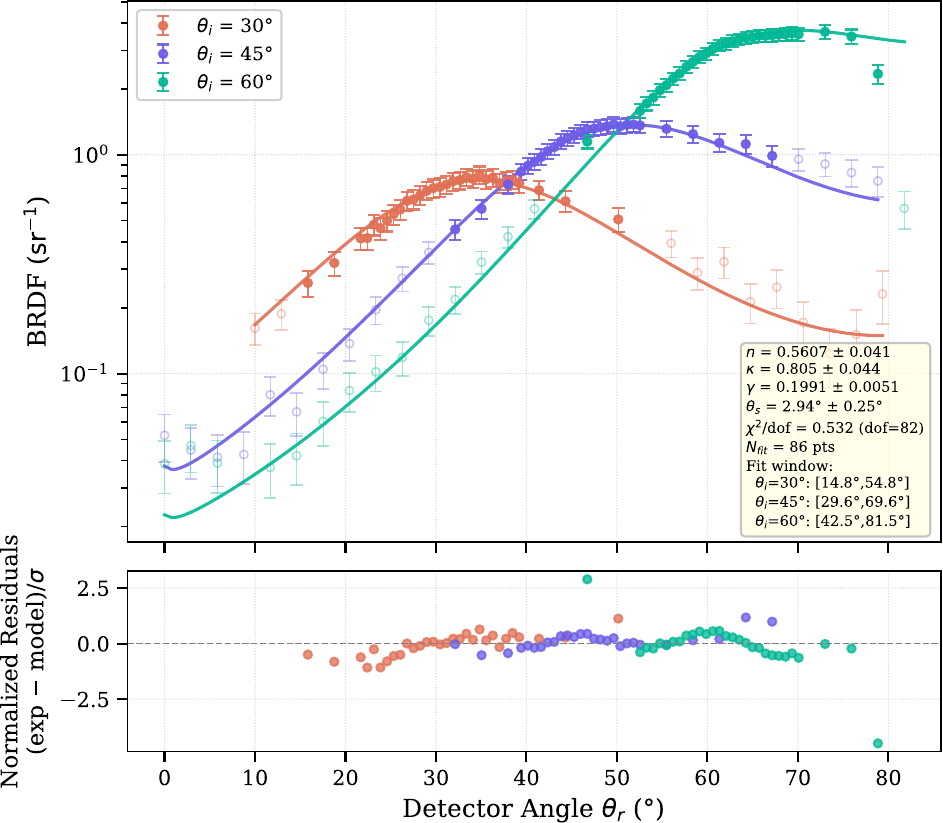}
            \caption{SS \qty{0.374}{\micro\meter}, \qty{175}{\nm}}
            \label{fig:brdf_fit_ss_175}
        \end{subfigure}
        \hfill
        \begin{subfigure}[b]{0.49\linewidth}
            \centering
            \includegraphics[width=\linewidth]{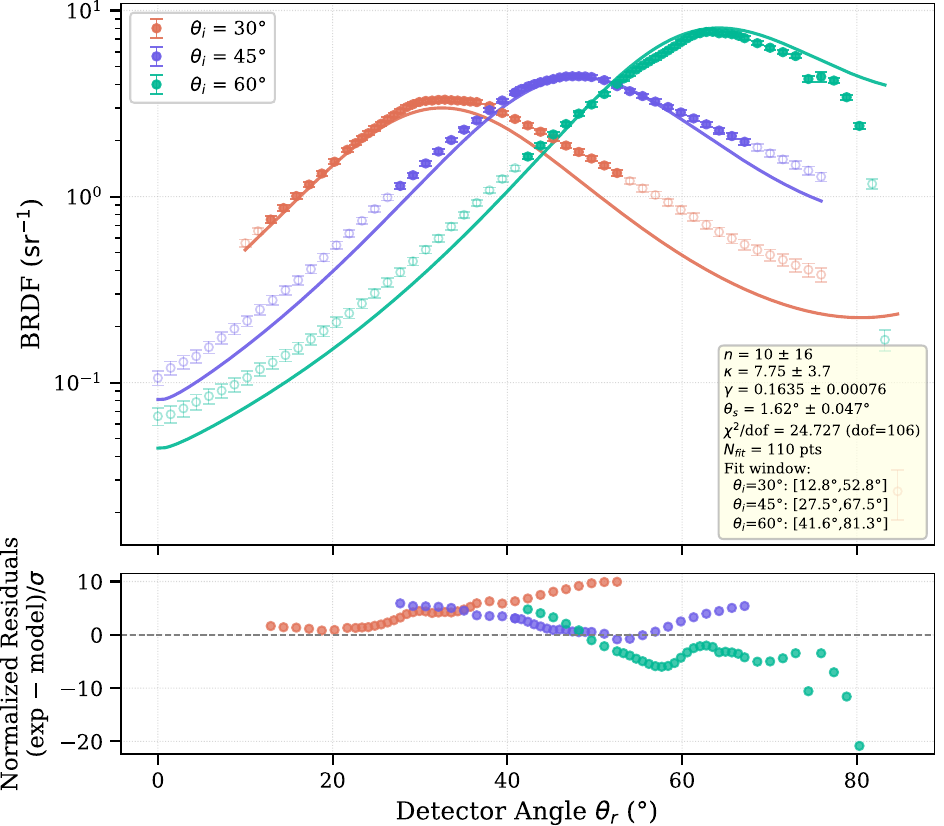}
            \caption{SS \qty{0.374}{\micro\meter}, \qty{500}{\nm}}
            \label{fig:brdf_fit_ss_500}
        \end{subfigure}
    \end{minipage}
    \caption{Experimental BRDF measurements (points) and corresponding GGX microfacet model fits (solid lines) for Al \qty{0.362}{\micro\meter} and SS \qty{0.374}{\micro\meter} at VUV (\qty{175}{\nm}) and visible (\qty{500}{\nm}) wavelengths on a semilogarithmic scale. Filled markers indicate points within the 20\% maximum BRDF threshold included in the fit, while empty markers indicate excluded points.}
    \label{fig:brdf_fits_results_4panel}
\end{figure*}

The angular BRDF scans for Al \qty{0.362}{\micro\meter} and SS \qty{0.374}{\micro\meter} are presented in \Cref{fig:brdf_fits_results_4panel} across three angles of incidence: \ang{30} (orange), \ang{45} (purple), and \ang{60} (teal). The BRDF profiles are shown on a semilogarithmic scale at \qty{175}{\nm} (VUV) and \qty{500}{\nm} (visible), while supplementary scans at \qtylist{128;150;200;300;400}{\nm} are included in \ref{subsec:app_fits_curves}. Only data points satisfying $\text{SNR} \ge 6$ are shown. For the \ang{30} AOI configuration, the angular scan is mechanically restricted to reflection angles $\theta_r \ge \ang{10}$ to prevent interference between the rotating photodetector assembly and the stationary motorized iris.

Across all configurations, the experimental BRDF profiles exhibit a prominent specular reflection lobe peaking near the direction ruled by Snell's law. In accordance with microfacet scattering theory, the specular lobe is asymmetric and shifted toward larger reflection angles relative to the nominal specular direction, falling off more sharply toward the surface normal than toward grazing angles. Said asymmetry originates from the angular variation of the Fresnel reflection coefficient at the microfacet, the shadowing-masking geometric attenuation, and the projected area factor $1/(\cos\theta_i \cos\theta_o)$. Moreover, the BRDF increases with the angle of incidence, driven by the angular rise of the Fresnel reflectance (\Cref{eq:fresnel_unpolarized}) and the reduced $\cos\theta_i$ factor in the denominator of \Cref{eq:BRDF_generic}.

At large reflection angles, particularly in the \ang{60} AOI scans, the measured BRDF values drop sharply near the grazing limit. This cutoff is attributed to geometric masking by the edges of the multi-sample holder assembly, which blocks a portion of the beam at extreme angles.

\begin{figure*}[!h]
    \centering
    \includegraphics[width=0.85\textwidth]{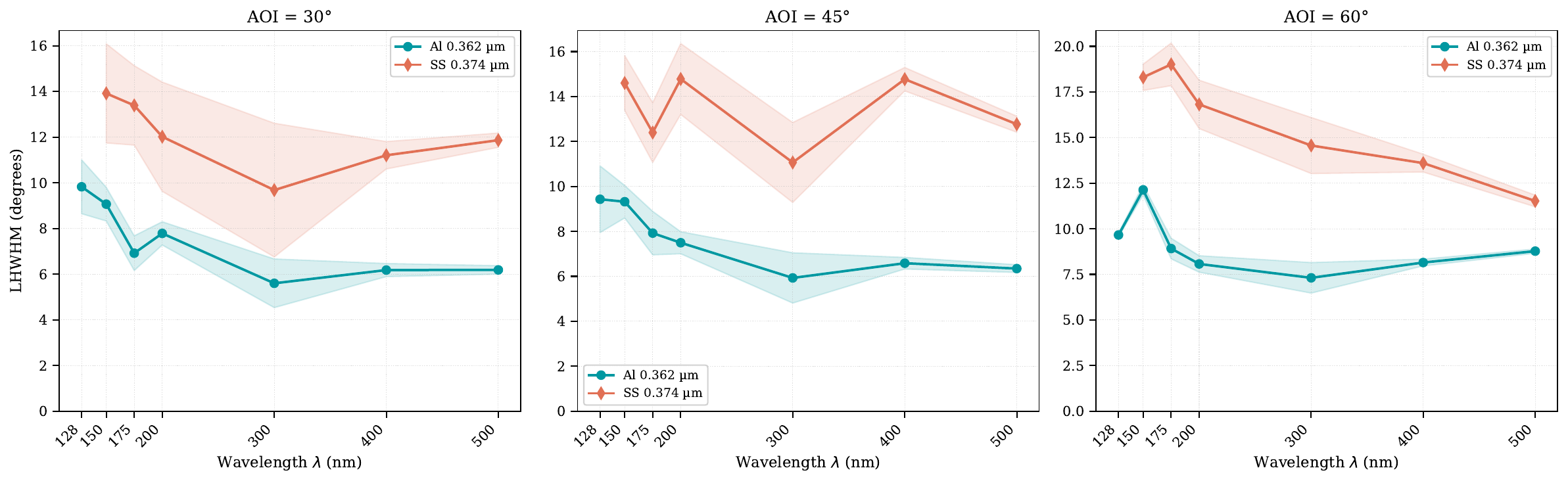}
    \caption{BRDF left half width at half maximum (LHWHM) as a function of wavelength for Al \qty{0.362}{\micro\meter} and SS \qty{0.374}{\micro\meter}.}
    \label{fig:hwhm_combined}
\end{figure*}

Since the right tail of the stainless steel BRDF is above the half maximum value at \ang{60} AOI (see \Cref{fig:brdf_fits_results_4panel}), the angular spread of the reflected beam is quantified by the specular lobe left half width at half maximum (LHWHM), evaluated from linear spline interpolation (\Cref{fig:hwhm_combined}). The stainless steel sample exhibits systematically broader specular peaks than the aluminum sample across the full spectral range, resulting in a more diffuse scattering behavior. The LHWHM generally exhibits a slight spectral widening toward shorter wavelengths for aluminum, except at \qtylist{128;500}{\nm} for \ang{60}. For stainless steel, this trend is similarly appreciated at \ang{60} AOI, yet the LHWHM presents no statistically significant trends at \ang{45} AOI. Larger error bands observed in \Cref{fig:hwhm_combined} for SS \qty{0.374}{\micro\meter} are expected because the BRDF profiles are flatter compared to Al \qty{0.362}{\micro\meter} angular distributions (see 
\Cref{fig:brdf_fits_results_4panel,fig:app_fits_al_semilog,fig:app_fits_ss_semilog}).

\begin{figure*}[!h]
    \centering
    \includegraphics[width=0.85\textwidth]{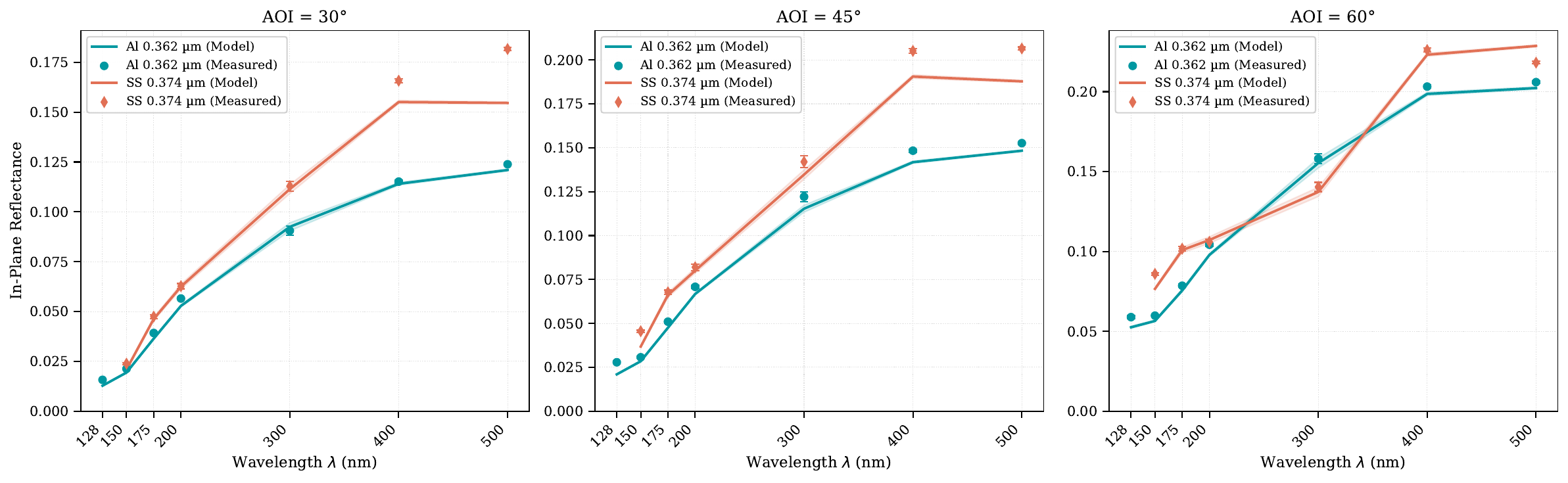}
    \caption{In-plane reflectance as a function of wavelength for Al \qty{0.362}{\micro\meter} and SS \qty{0.374}{\micro\meter}, comparing the model-independent spline integration (points with error bars) to the GGX model prediction (lines).}
    \label{fig:in_plane_reflectance}
\end{figure*}

The in-plane reflectance $R_{\text{in-plane}}$ is shown in \Cref{fig:in_plane_reflectance}, comparing model-independent numerical spline integration against the in-plane integral of the fitted GGX model, whose results are presented in \Cref{subsec:ggx_parameters}. Both metals exhibit a monotonic increase in reflectance with wavelength and angle of incidence, rising from approximately 2\% at \qty{128}{\nm} for \ang{30} AOI to above 20\% at \qty{500}{\nm} for \ang{60} AOI.

The model-independent and model-dependent values demonstrate consistent overall agreement. The empirical values are slightly higher than the single-scattering GGX predictions in the visible regime (notably at \qtylist{400;500}{\nm} for stainless steel) because direct integration captures multiple-scattered light in the BRDF tails, which the single-scattering GGX model neglects. At \qty{500}{\nm} and \ang{60} AOI for stainless steel, the model-independent reflectance exhibits a local suppression caused by sample holder masking of the elongated elliptical beam spot at large AOI. In contrast, the analytical GGX model does not reproduce this instrumental artifact.

\subsection{Optical Parameters of the GGX Microfacet Model}
\label{subsec:ggx_parameters}

Prior to extracting the hemispherical reflectance, the BRDF data is fitted to the GGX model via a multi-angular fit across \ang{30}, \ang{45}, and \ang{60} AOI of any given sample and wavelength combination. To isolate single-scattering microfacet reflections from multiple-scattering contributions, the fit window was restricted to experimental points exceeding 20\% of the maximum measured BRDF value (filled points in \Cref{fig:brdf_fits_results_4panel}). This threshold excludes the multiple-scattering tails that would distort the physical parameters of the GGX model. Additionally, grazing angle points which suffer from multi-sample holder masking are accordingly excluded from the fit.

The best-fit physical parameters ($n, \kappa, \gamma, \theta_s$), their associated $1\sigma$ uncertainties, and the corresponding reduced chi-squared values ($\chi^2/\text{ndf}$) extracted from multi-angular GGX BRDF fits are listed in \Cref{tab:brdf_parameter_fits}. 

\begin{table*}[!h]
    \centering
    \caption{Fitted parameters of the GGX BRDF model for Al \qty{0.362}{\micro\meter} and SS \qty{0.374}{\micro\meter}. The stainless steel measurement at \qty{128}{\nm} was excluded due to substantial gaseous argon purity drift.}
    \label{tab:brdf_parameter_fits}
    \small
    \begin{tabular}{l c c c S[table-format=2.3]}
        \toprule
        \multirow{2}{*}{Sample} & \multirow{2}{*}{$\lambda$ (\unit{\nm})} & Refractive Index                                                                                & Surface Roughness $\gamma$                                                                                       & {\multirow{2}{*}{$\chi^2/\text{ndf}$}} \\
                                &                                         & $\tilde{n} = n + i\kappa$                                                                       & Offset $\theta_s (^\circ)$                                                                                                & {}                                     \\
        \midrule
        Al \qty{0.362}{\micro\meter}
                                & 128                                     & \begin{tabular}[c]{@{}l@{}} $n = 1.10 \pm 0.02$ \\ $\kappa = 0.434 \pm 0.009$ \end{tabular}     & \begin{tabular}[c]{@{}l@{}} $\gamma = 0.123 \pm 0.003$ \\ $\theta_s = 1.89 \pm 0.12$ \end{tabular}   & 3.921                                  \\ \addlinespace
                                & 150                                     & \begin{tabular}[c]{@{}l@{}} $n = 1.41 \pm 0.03$ \\ $\kappa = 0.500 \pm 0.017$ \end{tabular}     & \begin{tabular}[c]{@{}l@{}} $\gamma = 0.1250 \pm 0.0015$ \\ $\theta_s = 1.03 \pm 0.09$ \end{tabular} & 2.851                                  \\ \addlinespace
                                & 175                                     & \begin{tabular}[c]{@{}l@{}} $n = 2.08 \pm 0.11$ \\ $\kappa = 0.00 \pm 83.79$ \end{tabular}      & \begin{tabular}[c]{@{}l@{}} $\gamma = 0.1146 \pm 0.0019$ \\ $\theta_s = 1.35 \pm 0.11$ \end{tabular} & 0.479                                  \\ \addlinespace
                                & 200                                     & \begin{tabular}[c]{@{}l@{}} $n = 2.16 \pm 0.10$ \\ $\kappa = 0.63 \pm 0.12$ \end{tabular}       & \begin{tabular}[c]{@{}l@{}} $\gamma = 0.1073 \pm 0.0013$ \\ $\theta_s = 1.84 \pm 0.07$ \end{tabular} & 1.791                                  \\ \addlinespace
                                & 300                                     & \begin{tabular}[c]{@{}l@{}} $n = 1.28 \pm 0.16$ \\ $\kappa = 1.41 \pm 0.08$ \end{tabular}       & \begin{tabular}[c]{@{}l@{}} $\gamma = 0.1022 \pm 0.0024$ \\ $\theta_s = 2.48 \pm 0.13$ \end{tabular} & 0.234                                  \\ \addlinespace
                                & 400                                     & \begin{tabular}[c]{@{}l@{}} $n = 0.97 \pm 0.03$ \\ $\kappa = 1.39 \pm 0.03$ \end{tabular}       & \begin{tabular}[c]{@{}l@{}} $\gamma = 0.0972 \pm 0.0006$ \\ $\theta_s = 2.38 \pm 0.04$ \end{tabular} & 1.325                                  \\ \addlinespace
                                & 500                                     & \begin{tabular}[c]{@{}l@{}} $n = 1.28 \pm 0.04$ \\ $\kappa = 1.59 \pm 0.02$ \end{tabular}       & \begin{tabular}[c]{@{}l@{}} $\gamma = 0.0927 \pm 0.0004$ \\ $\theta_s = 1.76 \pm 0.02$ \end{tabular} & 6.445                                  \\
        \midrule
        SS \qty{0.374}{\micro\meter}
                                & 150                                     & \begin{tabular}[c]{@{}l@{}} $n = 0.79 \pm 0.01$ \\ $\kappa = 0.52 \pm 0.01$ \end{tabular}       & \begin{tabular}[c]{@{}l@{}} $\gamma = 0.165 \pm 0.003$ \\ $\theta_s = 1.91 \pm 0.14$ \end{tabular}   & 8.872                                  \\ \addlinespace
                                & 175                                     & \begin{tabular}[c]{@{}l@{}} $n = 0.56 \pm 0.04$ \\ $\kappa = 0.81 \pm 0.04$ \end{tabular}       & \begin{tabular}[c]{@{}l@{}} $\gamma = 0.199 \pm 0.005$ \\ $\theta_s = 2.94 \pm 0.25$ \end{tabular}   & 0.532                                  \\ \addlinespace
                                & 200                                     & \begin{tabular}[c]{@{}l@{}} $n = 0.75 \pm 0.09$ \\ $\kappa = 1.33 \pm 0.09$ \end{tabular}       & \begin{tabular}[c]{@{}l@{}} $\gamma = 0.192 \pm 0.005$ \\ $\theta_s = 3.65 \pm 0.27$ \end{tabular}   & 0.546                                  \\ \addlinespace
                                & 300                                     & \begin{tabular}[c]{@{}l@{}} $n = 8.98 \pm 31.20$ \\ $\kappa = 0.00 \pm 137734.00$ \end{tabular} & \begin{tabular}[c]{@{}l@{}} $\gamma = 0.181 \pm 0.005$ \\ $\theta_s = 3.09 \pm 0.27$ \end{tabular}   & 0.330                                  \\ \addlinespace
                                & 400                                     & \begin{tabular}[c]{@{}l@{}} $n = 7.42 \pm 20.41$ \\ $\kappa = 10.00 \pm 6.63$ \end{tabular}     & \begin{tabular}[c]{@{}l@{}} $\gamma = 0.1703 \pm 0.0011$ \\ $\theta_s = 2.62 \pm 0.07$ \end{tabular} & 6.904                                  \\ \addlinespace
                                & 500                                     & \begin{tabular}[c]{@{}l@{}} $n = 10.00 \pm 15.50$ \\ $\kappa = 7.75 \pm 3.70$ \end{tabular}     & \begin{tabular}[c]{@{}l@{}} $\gamma = 0.1635 \pm 0.0008$ \\ $\theta_s = 1.62 \pm 0.05$ \end{tabular} & 24.727                                 \\
        \bottomrule
    \end{tabular}
\end{table*}

The reduced chi-squared values for aluminum generally range between $\chi^2/\text{ndf} \approx 0.2$ and \num{3.9}, except at \qty{500}{\nm} where it reaches $\chi^2/\text{ndf} \approx 6.4$. For stainless steel, reduced chi-squared values are similar in the VUV, rising to \num{24.7} at \qty{500}{\nm}. These elevated $\chi^2/\text{ndf}$ values identified in specific datasets, particularly at \qty{500}{\nm} for stainless steel, point to potential limitations in either the evaluation of the experimental systematic uncertainties or the single-scattering GGX description, or both. An example of experimental systematic deviations is the geometric masking due to the multi-sample holder edges, while multiple-scattering contributions in the BRDF tails are a known limitation of the BRDF model. These larger $\chi^2/\text{ndf}$ may partially reflect the very small experimental uncertainties achieved at high lamp intensities in the visible spectrum, which can penalize minor residual deviations between the fitted curves and the data. 

Aside from the reduced chi-squared metric, the fitted geometric parameters in~\Cref{tab:brdf_parameter_fits} display consistent physical trends for both materials. The roughness $\gamma$ is the dominant parameter, controlling the angular width and shape of the specular lobe. For the aluminum (stainless steel) sample, $\gamma$ lies within the range \numrange{0.093}{0.125} (\numrange{0.164}{0.199}). This reflects the broader specular lobes measured for the stainless steel sample, quantified in \Cref{fig:hwhm_combined}. In addition, both metals exhibit a slight decrease in $\gamma$ toward longer wavelengths, again consistently with \Cref{fig:hwhm_combined} trends. As the microfacet distribution is an intrinsic geometric feature of the surface, this slight wavelength dependence suggests that $\gamma$ may act as an effective parameter partially absorbing wave-optics effects at long wavelengths, where the wavelength can be comparable to the root-mean-square roughness \cite{Beckmann1963}. On the other hand, the angular offset $\theta_s$ ranges between \ang{1.0} and \ang{2.5} (\ang{1.6} and \ang{3.7}) for the aluminum (stainless steel) sample, quantifying the accuracy of our angular alignment procedure.

The $1\sigma$ uncertainties in~\Cref{tab:brdf_parameter_fits} reveal distinct behaviors between the geometric parameters ($\gamma, \theta_s$) and the optical parameters ($n, \kappa$). $\gamma$ and $\theta_s$ are constrained across all wavelengths, with relative uncertainties typically below 3\% for $\gamma$ and angular uncertainties below \ang{0.3}, respectively. In contrast, the optical parameters exhibit considerably large uncertainties in specific configurations. Particularly, for the stainless steel sample at \qtylist{300;400;500}{\nm}, as well as for the aluminum sample at \qty{175}{\nm}, the fit converges toward the preset parameter boundaries ($\kappa \to 0$ or $n \to 10$), accompanied by very large fit uncertainties (e.g., $n = 10.00 \pm 15.50$ for SS at \qty{500}{\nm}). While boundary convergences may suggest numerical or modeling mismatching, in our case they are largely driven by an intrinsic, physical degeneracy between the real ($n$) and imaginary ($\kappa$) components of the complex refractive index. Such degeneracy is studied by means of the correlation matrices.

\begin{figure*}[!h]
    \centering
    \begin{minipage}{0.75\textwidth}
        \begin{subfigure}[b]{0.49\linewidth}
            \centering
            \includegraphics[width=\linewidth]{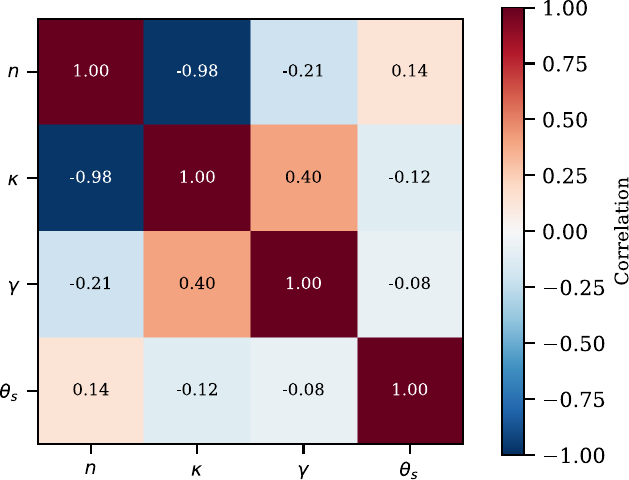}
            \caption{Al \qty{0.362}{\micro\meter}, \qty{175}{\nm}}
            \label{fig:corr_al_175}
        \end{subfigure}
        \hfill
        \begin{subfigure}[b]{0.49\linewidth}
            \centering
            \includegraphics[width=\linewidth]{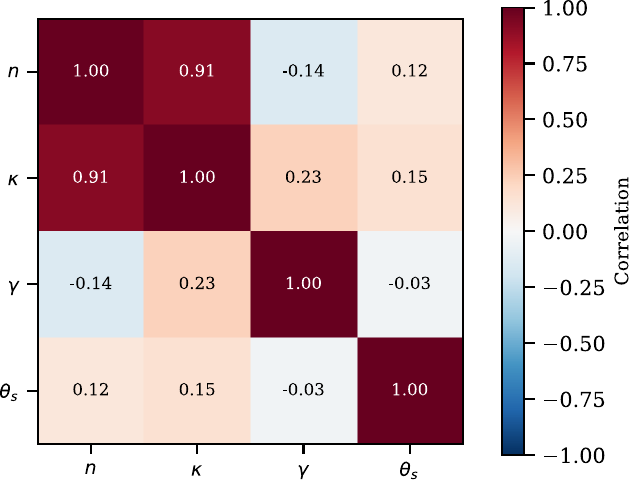}
            \caption{Al \qty{0.362}{\micro\meter}, \qty{500}{\nm}}
            \label{fig:corr_al_500}
        \end{subfigure}
    
        \vspace{0.2cm}
    
        \begin{subfigure}[b]{0.49\linewidth}
            \centering
            \includegraphics[width=\linewidth]{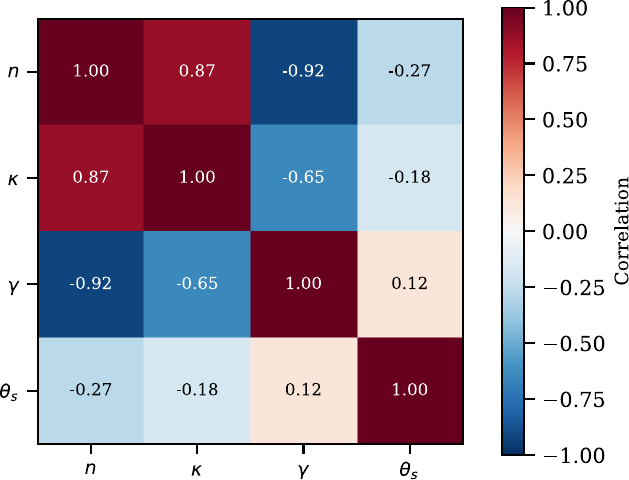}
            \caption{SS \qty{0.374}{\micro\meter}, \qty{175}{\nm}}
            \label{fig:corr_ss_175}
        \end{subfigure}
        \hfill
        \begin{subfigure}[b]{0.49\linewidth}
            \centering
            \includegraphics[width=\linewidth]{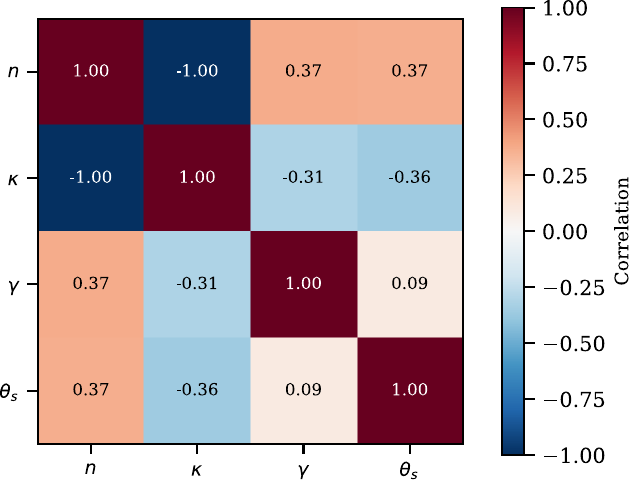}
            \caption{SS \qty{0.374}{\micro\meter}, \qty{500}{\nm}}
            \label{fig:corr_ss_500}
        \end{subfigure}
    \end{minipage}
    \caption{Correlation matrices between the fitted parameters ($n$, $\kappa$, $\gamma$, $\theta_s$) for Al \qty{0.362}{\micro\meter} and SS \qty{0.374}{\micro\meter} at \qty{175}{\nm} (VUV) and \qty{500}{\nm} (visible) wavelengths.}
    \label{fig:corr_matrices_4panel}
\end{figure*}

The correlation matrices of the fits are shown in \Cref{fig:corr_matrices_4panel} at \qtylist{175;500}{\nm}, while for other wavelengths they are provided in \ref{sec:app_correlations}.

A pronounced correlation between the two components of the refractive index is observed. Measuring unpolarized light constrains the magnitude of the unpolarized Fresnel coefficient (\Cref{eq:fresnel_unpolarized}), rather than the individual polarization components $R_s$ and $R_p$. Consequently, multiple combinations of $(n, \kappa)$ can reproduce compatible unpolarized Fresnel reflection curves, creating a degenerate $\chi^2$ valley in the $(n, \kappa)$ parameter space. Moreover, the sign of the correlation between $n$ and $\kappa$ can be understood by the analytical structure of the Fresnel equations and the value of the real and imaginary parts of the refractive index. 
The roughness $\gamma$ and alignment offset $\theta_s$ remain decoupled from the optical parameters $n$ and $\kappa$ for the majority of the wavelengths; however, this is not the case for some measurements. For example, the stainless steel sample at \qty{175}{\nm} exhibits a strong correlation of \num{-0.92} between $\gamma$ and $n$, and \num{-0.65} between $\gamma$ and $\kappa$, suggesting that $\gamma$ and the Fresnel coefficient can become correlated in specific cases.

Although this degeneracy implies that the individual best-fit values of $n$ and $\kappa$ should not be interpreted as accurate optical parameters of the sample, it does not compromise the reconstruction of the BRDF profile or the integrated hemispherical reflectance. Since the fit optimizer moves along the valley of approximately constant $F(\theta'; n, \kappa)$, the reflectance is reproduced, while the spatial shape and angular lobe width remain primarily constrained by the roughness parameter $\gamma$.

\subsection{Hemispherical Reflectance}
\label{subsec:hemispherical_reflectance}

\begin{figure*}[!h]
    \centering
    \begin{minipage}{\textwidth}
        \centering
        \begin{subfigure}[b]{0.49\linewidth}
            \centering
            \includegraphics[width=\linewidth]{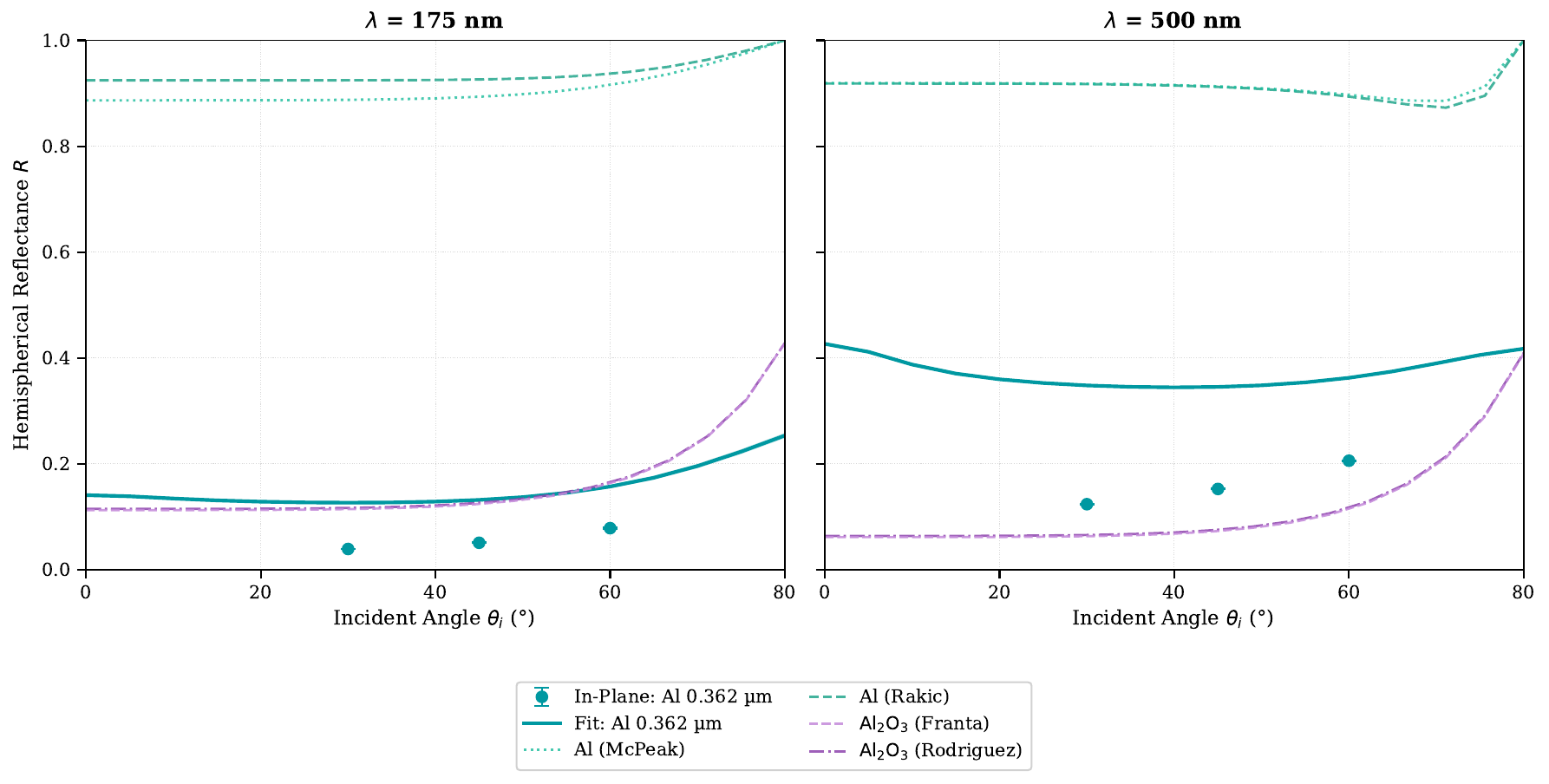}
            \caption{Al \qty{0.362}{\micro\meter} comparison with literature benchmarks}
            \label{fig:hemi_aoi_al_benchmark}
        \end{subfigure}
        \hfill
        \begin{subfigure}[b]{0.49\linewidth}
            \centering
            \includegraphics[width=\linewidth, trim=0 -16.328bp 0 0]{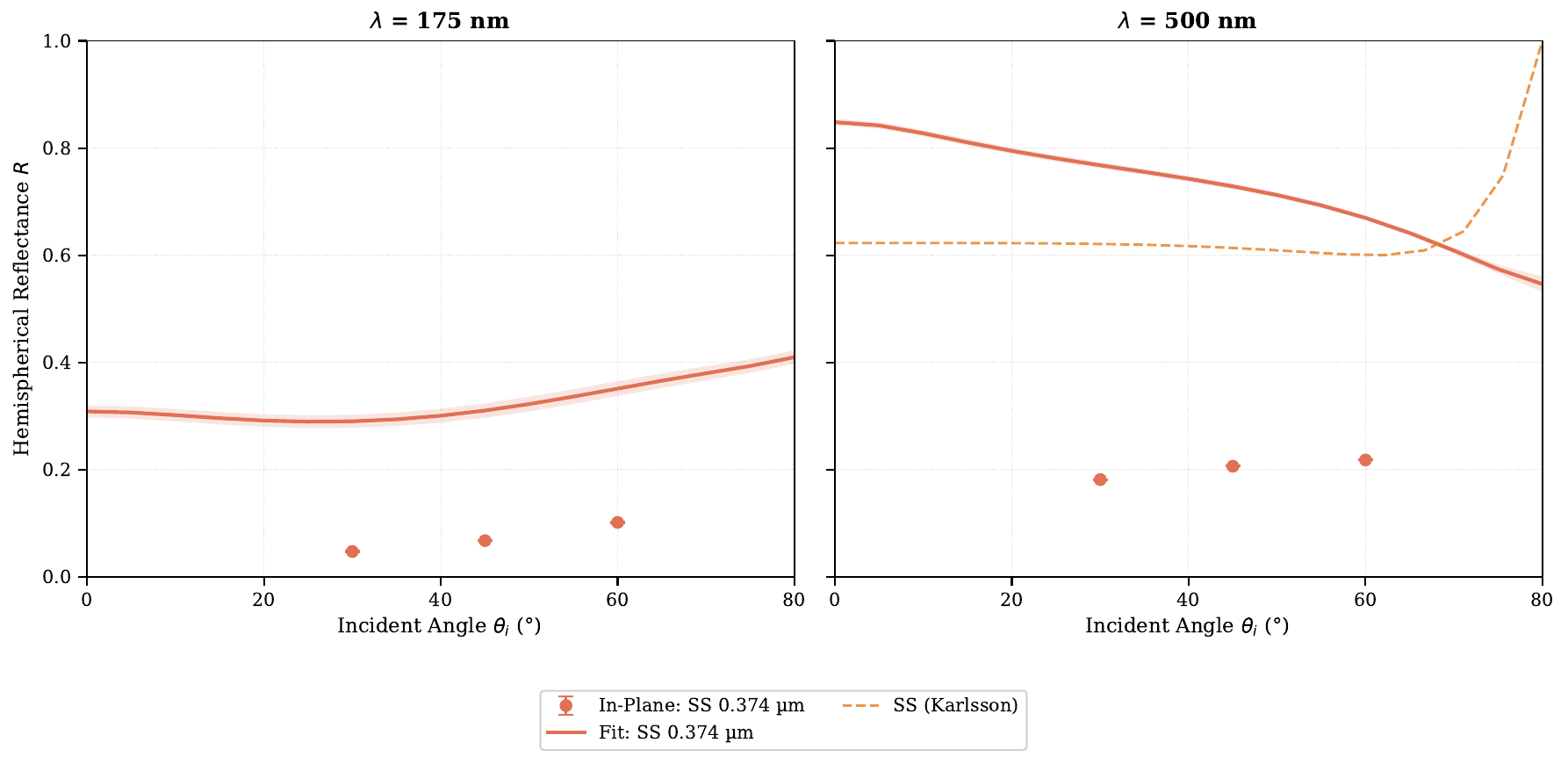}
            \caption{SS \qty{0.374}{\micro\meter} comparison with literature benchmarks}
            \label{fig:hemi_aoi_ss_benchmark}
        \end{subfigure}
    \end{minipage}
    \caption{Integrated hemispherical reflectance as a function of the angle of incidence for Al \qty{0.362}{\micro\meter} and SS \qty{0.374}{\micro\meter} compared with literature benchmarks (dashed curves) and model-independent in-plane reflectances (points).}
    \label{fig:hemi_aoi_comparison}
\end{figure*}

The total hemispherical reflectance is determined by integrating the fitted GGX BRDF model over the hemisphere. The resulting hemispherical reflectance curves as a function of angle of incidence are depicted at \qtylist{175;500}{\nm} in \Cref{fig:hemi_aoi_comparison}, together with the model-independent in-plane reflectances (points). The remaining wavelengths are available in \Cref{fig:app_hemi_refl_vs_aoi}. The discussion relative to the literature curves (dashed lines) is conducted in \Cref{sec:discussion}.

The dependence of hemispherical reflectance on the angle of incidence reveals two distinct regimes. In the VUV range ($\lambda \le \qty{200}{\nm}$), the reflectance increases monotonically with the angle of incidence, driven by the angular rise of the Fresnel reflection coefficient toward grazing angles. At longer wavelengths ($\lambda \gtrsim \qty{300}{\nm}$), the angular dependence flattens or decreases with the angle of incidence. In this regime, the microfacet shadowing-masking factor ($G_2$ in \Cref{eq:BRDF_generic,eq:masking-shadowing-g2}, decreasing with AOI) counterbalances the Fresnel coefficient gain.

\begin{figure*}[!h]
    \centering
    \begin{minipage}{0.85\textwidth}
        \begin{subfigure}[b]{0.49\textwidth}
            \centering
            \includegraphics[width=\linewidth]{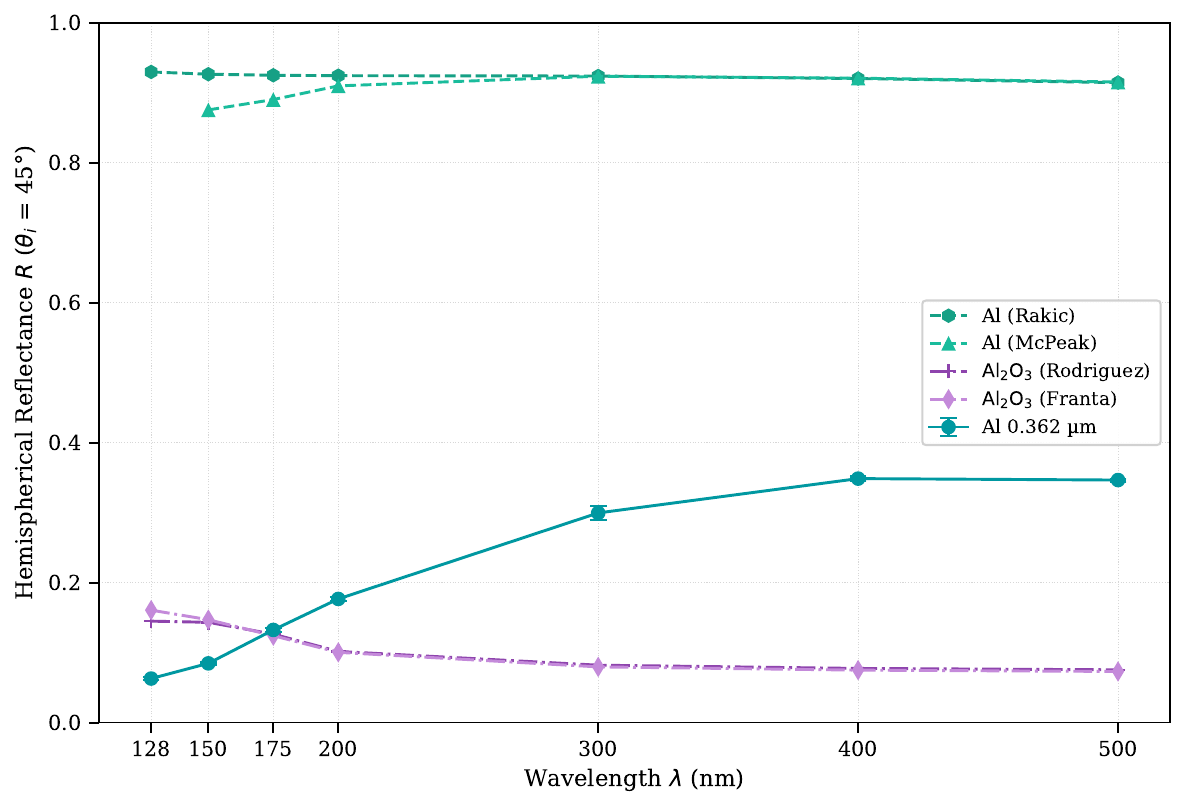}
            \caption{Al \qty{0.362}{\micro\meter} comparison with \ce{Al2O3}}
            \label{fig:hemi_wave_al_al2o3}
        \end{subfigure}
        \hfill
        \begin{subfigure}[b]{0.49\textwidth}
            \centering
            \includegraphics[width=\linewidth]{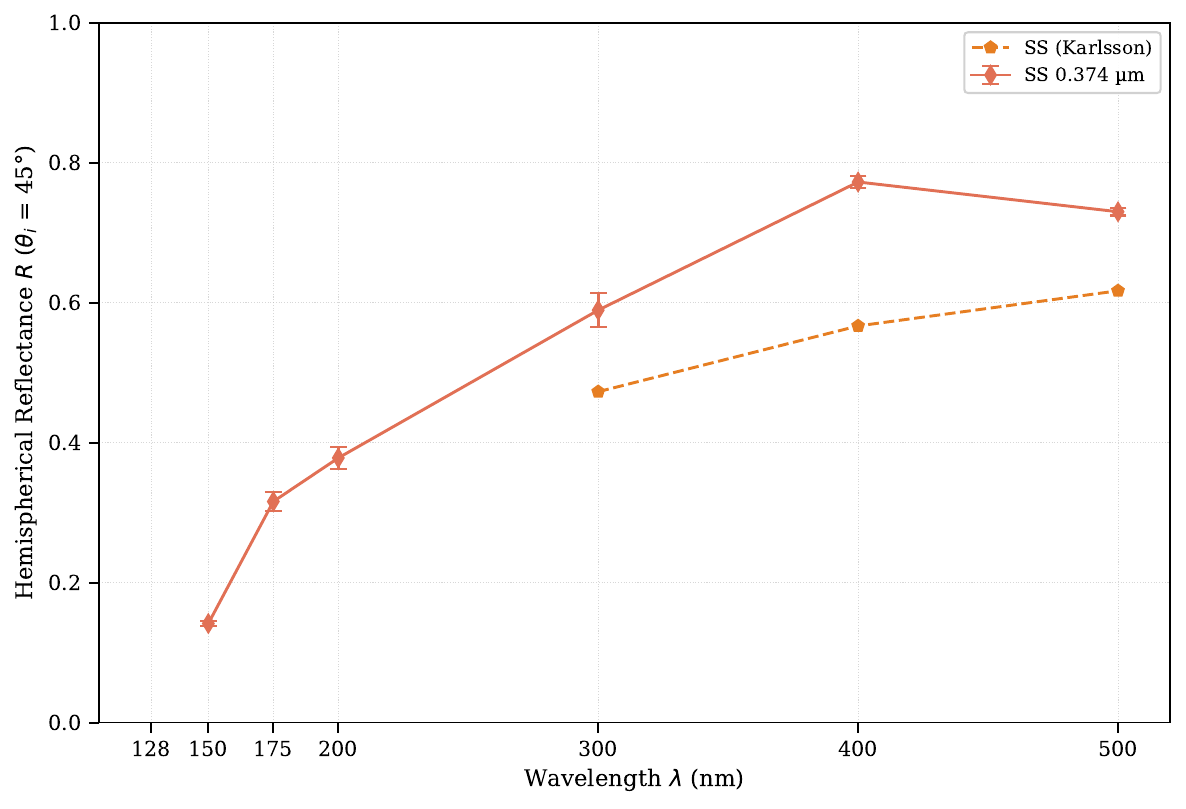}
            \caption{SS \qty{0.374}{\micro\meter} comparison}
            \label{fig:hemi_wave_ss}
        \end{subfigure}
    \end{minipage}
    \caption{Spectral comparison of the integrated hemispherical reflectance at \ang{45} AOI for Al \qty{0.362}{\micro\meter} and SS \qty{0.374}{\micro\meter}. Literature curves for \ce{Al2O3} are superimposed on the aluminum results (a) to assess native oxide absorption effects.}
    \label{fig:hemi_wavelength_comparison_al2o3}
\end{figure*}

The spectral dependence of the hemispherical reflectance at a reference AOI of \ang{45} is presented in \Cref{fig:hemi_wavelength_comparison_al2o3}, compared against literature benchmarks.

For the aluminum sample, the hemispherical reflectance drops sharply below \qty{200}{\nm} ($R < 20\%$), while the maximum reflectance approaches 40\% at \qty{400}{\nm}. For the stainless steel sample, the hemispherical reflectance varies between 15\% and 40\% in the VUV, rising across the visible spectrum, with highest reflectance nearing 80\% at \qty{400}{\nm}. A physical interpretation of these spectral features, as well as the comparison with the literature benchmarks, are developed in \Cref{sec:discussion}.
\section{Discussion} \label{sec:discussion}

\subsection{Comparison with Literature Benchmarks}
\label{subsec:discussion_literature}

A primary finding of this analysis is the notable suppression of the hemispherical reflectance of the Al \qty{0.362}{\micro\meter} compared to standard thin-film aluminum mirror benchmarks (teal dashed lines in \Cref{fig:hemi_aoi_al_benchmark,fig:hemi_wave_al_al2o3}), which report outstanding reflectance properties across the experimental wavelengths~\cite{Rakic1995, McPeak2015}. This discrepancy indicates that the reflective performance of the reference aluminum sample cannot be described by a clean aluminum surface. In contrast, the spontaneous formation of a native oxide (\ce{Al2O3}) layer on bare aluminum by ambient passivation is expected. A thin oxide layer on a metallic surface introduces multilayer interference and light absorption, affecting the reflective properties. Literature benchmarks for thin \ce{Al2O3} layers (purple dashed lines)~\cite{Franta2015, Rodriguez2025} suggest that said hypothesis is responsible for the observed lower reflectance for the aluminum sample.

For SS \qty{0.374}{\micro\meter}, the literature benchmark corresponds to highly polished 304 stainless steel~\cite{Karlsson1982} (orange dashed lines in (\Cref{fig:hemi_wave_ss,fig:hemi_wave_ss})), only available in the UV-VIS range. In this region, the hemispherical reflectance measured for our reference stainless steel sample moderately exceeds the literature benchmark, except at near grazing angles, as discussed in \Cref{sec:results}. The increased reflectance with respect to literature values is plausibly attributed to variations in surface passivation and finishing, alongside potential minor overestimation in the out-of-plane scattering extrapolation of the GGX model for broad lobes. Given the lack of VUV optical data for stainless steel 304, these measurements provide a valuable empirical reference for rough stainless steel surfaces relevant to detector materials.

\subsection{Impact of Detector Media on the Reflectance}
\label{subsec:detector_media}

In cryogenic noble-element detectors, detector materials are typically immersed in liquid argon or liquid xenon. According to the microfacet theory, modifying the dielectric immersion medium from gaseous argon or xenon ($n_0 \approx 1.0$) to a liquid phase alters the electromagnetic boundary conditions through the refractive index of the immersion medium $n_0(\lambda)$.

We have computed $n_0(\lambda)$ for LAr and LXe using the Sellmeier relation~\cite{Sellmeier2017}:
\begin{equation}
    \label{eq:sellmeier_liquid_nobles}
    n_0^2(\lambda) = A_0 + \frac{A_{\mathrm{UV}} \lambda^2}{\lambda^2 - \lambda_{\mathrm{UV}}^2} + \frac{A_{\mathrm{IR}} \lambda^2}{\lambda^2 - \lambda_{\mathrm{IR}}^2},
\end{equation}
with parameters for LAr at \qty{90}{\K} ($A_0 = 1.26 \pm 0.09$, $A_{\mathrm{UV}} = 0.23 \pm 0.09$, $\lambda_{\mathrm{UV}} = \qty{106.6}{\nm}$, $A_{\mathrm{IR}} = 0.0023 \pm 0.007$, $\lambda_{\mathrm{IR}} = \qty{908.3}{\nm}$) and LXe at \qty{178}{\K} ($A_0 = 1.4 \pm 0.2$, $A_{\mathrm{UV}} = 0.4 \pm 0.2$, $\lambda_{\mathrm{UV}} = \qty{146.9}{\nm}$, $A_{\mathrm{IR}} = 0.002 \pm 0.02$, $\lambda_{\mathrm{IR}} = \qty{827.0}{\nm}$)~\cite{Sellmeier2017, Sinnock1969}. For liquid xenon, values below \qty{175}{\nm} are omitted because the medium becomes opaque near its absorption resonance at $\lambda_{\mathrm{UV}} = \qty{146.9}{\nm}$.

\begin{figure*}[!h]
    \centering
    \includegraphics[width=0.85\textwidth]{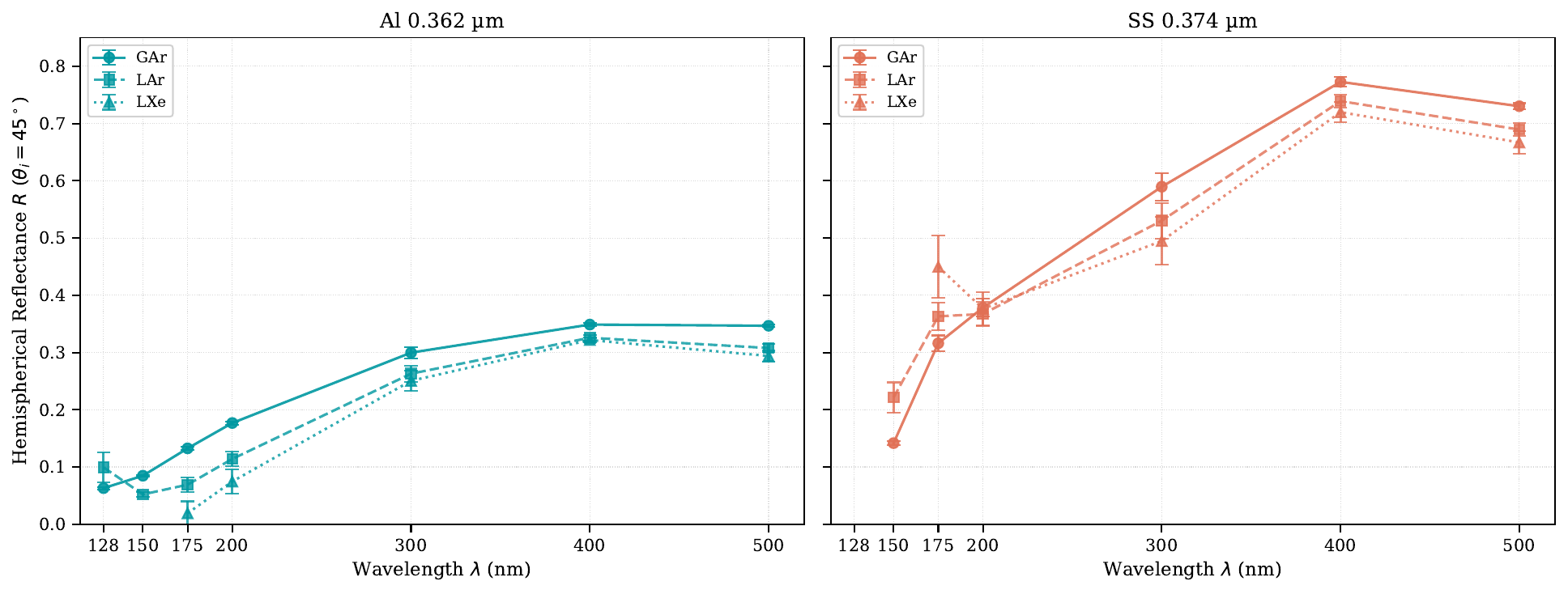}
    \caption{Extrapolated hemispherical reflectance at \ang{45} AOI for Al \qty{0.362}{\micro\meter} (left) and SS \qty{0.374}{\micro\meter} (right) in gaseous argon compared to predictions in liquid argon and liquid xenon using the GGX microfacet model. Error bars denote $1\sigma$ uncertainties propagated from parameter covariance and medium refractive index, summed in quadrature. Liquid xenon values below \qty{175}{\nm} are omitted due to absorption.}
    \label{fig:hemi_reflectance_liquid_nobles_ggx}
\end{figure*}

From the refractive indices, the hemispherical reflectance is extrapolated leveraging the GGX model. In \Cref{fig:hemi_reflectance_liquid_nobles_ggx}, the calculations at \ang{45} AOI in gaseous argon (equivalent to gaseous xenon), liquid argon, and liquid xenon are compared.

In the UV-VIS region, immersion in a denser dielectric medium suppresses the sample reflectance between \qtyrange{4}{12}{\%} for LAr and and \qtyrange{7}{16}{\%} LXe, following the hierarchy $R_{\text{GAr}} > R_{\text{LAr}} > R_{\text{LXe}}$. However, an inversion of the pattern arises in the VUV. For SS \qty{0.374}{\micro\meter}, the hemispherical reflectance for the three media converge around at \qty{200}{\nm}. At \qtylist{150;175}{\nm}, the reflectance in the liquid phase exceeds that in GAr. Additionally, the reflectance for LXe is higher than LAr at \qtylist{175;200}{\nm}. For Al \qty{0.362}{\micro\meter}, this inversion is observed at \qty{128}{\nm} between LAr and GAr. This phenomenon emerges from the Fresnel equations at conductive boundaries, explained in \Cref{subsec:fresnel_reflections}, involving the medium and sample refractive indices, as well as the angle of incidence.
\section{Conclusions}
\label{sec:conclusions}

To determine the reflective properties of materials used in noble-element particle detectors, an automated rotary setup was developed, enabling angle-resolved reflectance measurements via the bidirectional reflectance distribution function (BRDF) in the VUV and UV-VIS wavelength ranges. Such measurements are conducted inside a dark box purged with high-purity gaseous argon at \qty{0.05}{bar} overpressure. Using this setup, two rough metallic samples were characterized at \ang{30}, \ang{45}, and \ang{60} AOI: aluminum AW-1050 (A), with \qty{0.362}{\micro\meter} average roughness, and stainless steel SS 304, with \qty{0.374}{\micro\meter} average roughness.

A model-independent analysis of the BRDF data within the specular reflection plane shows that the specular lobe width is slightly wavelength dependent, widening toward shorter wavelengths, which might suggest the appearance of wave-optics effects in the VUV. On the other hand, the in-plane reflectance increases monotonically with wavelength and AOI for both metals, ranging from approximately \qty{2}{\percent} for \qty{128}{\nm} at \ang{30} AOI to above \qty{20}{\percent} for \qty{500}{\nm} at \ang{60} AOI.

The fit of the BRDF profiles via the GGX microfacet model constrains the roughness parameter to \numrange{0.093}{0.125} for the aluminum sample and \numrange{0.164}{0.199} for the stainless steel, compatibly with the broader specular lobe observed for the latter. Despite the strong correlation between the real refractive index and the extinction coefficient induced by unpolarized light, the Fresnel reflection coefficient is reliably inferred.

The integration over the hemisphere of the BRDF described by the fitted GGX model in gaseous argon (and likewise for other gaseous media) provides the hemispherical reflectance as a function of the angle of incidence across the experimental wavelengths. Additionally, the hemispherical reflectance is studied as a function of wavelength at the reference angle of \ang{45} AOI. The reflectance for the aluminum sample remains below \qty{20}{\percent} for wavelengths shorter than \qty{200}{\nm} and approaches \qty{40}{\percent} in the visible range. These values can be justified by an oxide layer on the bare sample surface. In contrast, for the stainless steel, the hemispherical reflectance at \ang{45} AOI ranges between \qtyrange{15}{40}{\percent} in the VUV and reaches nearly \qty{80}{\percent} at \qty{400}{\nm}. According to \cite{polyanskiy2024}, these measurements could be the first characterization of the reflectance for SS 304 in the VUV.

At last, the hemispherical reflectance is extrapolated to liquid argon and liquid xenon at \ang{45} AOI, where a change of immersion medium exhibits a change of the reflectance behavior via the Fresnel reflection coefficient, particularly in the VUV.

Having been validated on reference metallic samples, the experimental setup and measurement methodology developed in this work provide a basis for characterizing candidate detector materials in dedicated follow-up campaigns.

\section*{CRediT authorship contribution statement}
\textbf{H~Amar:} Conceptualization, Methodology, Software, Formal analysis, Investigation, Data Curation, Writing - Original Draft, Visualization. \textbf{J.~\'Alvarez:} Methodology, Resources. \textbf{X.~Arrue:} Investigation. \textbf{C.~Benitez:} Methodology, Resources, Software. \textbf{A.~Cervera-Villanueva:} Conceptualization, Supervision, Funding acquisition. \textbf{J.~Mart\'in-Albo:} Conceptualization, Supervision, Funding acquisition. \textbf{L.~Molina-Bueno:} Supervision. \textbf{A.~Roche:} Investigation. \textbf{M.~Sorel:} Writing - Original Draft, Supervision, Funding acquisition. \textbf{J.~Soto-Oton:} Conceptualization, Methodology, Software, Investigation, Data Curation, Supervision. \textbf{S.~Teruel-Pardo:} Methodology, Resources. \textbf{N.~Yahlali:} Supervision, Funding acquisition.

\section*{Declaration of competing interest}
The authors declare that they have no known competing financial interests or personal relationships that could have appeared to influence the work reported in this paper.

\section*{Acknowledgments}
This work was supported by the Spanish \textit{Ministerio de Ciencia, Innovación y Universidades} (MICIU), the \textit{Agencia Estatal de Investigaci\'on} (AEI) under grant agreement No. PID2023-147949NB-C52 / 10.13039/501100011033, and co-funded by the European Union (FEDER). Additionally, the authors acknowledge the financial support from the \textit{Ministerio de Ciencia e Innovación} (MICIN) with funding from the European Union NextGenerationEU (PRTR-C17.I01) and Generalitat Valenciana under the project IFIC-NuLight, ASFAE/2022/029.

\section*{Data availability}
Data will be made available on request.

\appendix
\section{Supplementary BRDF Fits and Hemispherical Reflectance}
\label{sec:app_fits}

This appendix compiles the supplementary fitted BRDF profiles on a semilogarithmic scale and the corresponding integrated hemispherical reflectance curves as a function of the angle of incidence. The fitting is performed using the GGX microfacet model for both the aluminum (AW-1050 (A), \qty{0.362}{\micro\meter} average roughness) and stainless steel (SS 304, \qty{0.374}{\micro\meter} average roughness) samples across the remaining analyzed wavelengths (\qtylist[list-units = single]{128;150;200;300;400}{\nm}). The stainless steel measurement at \qty{128}{\nm} was omitted due to substantial gaseous argon purity drift.

\subsection{Fitted BRDF Curves}
\label{subsec:app_fits_curves}

\begin{figure*}[!htbp]
    \centering
    \begin{minipage}{0.9\textwidth}
        \centering
        \begin{subfigure}[b]{0.32\linewidth}
            \centering
            \includegraphics[width=\linewidth]{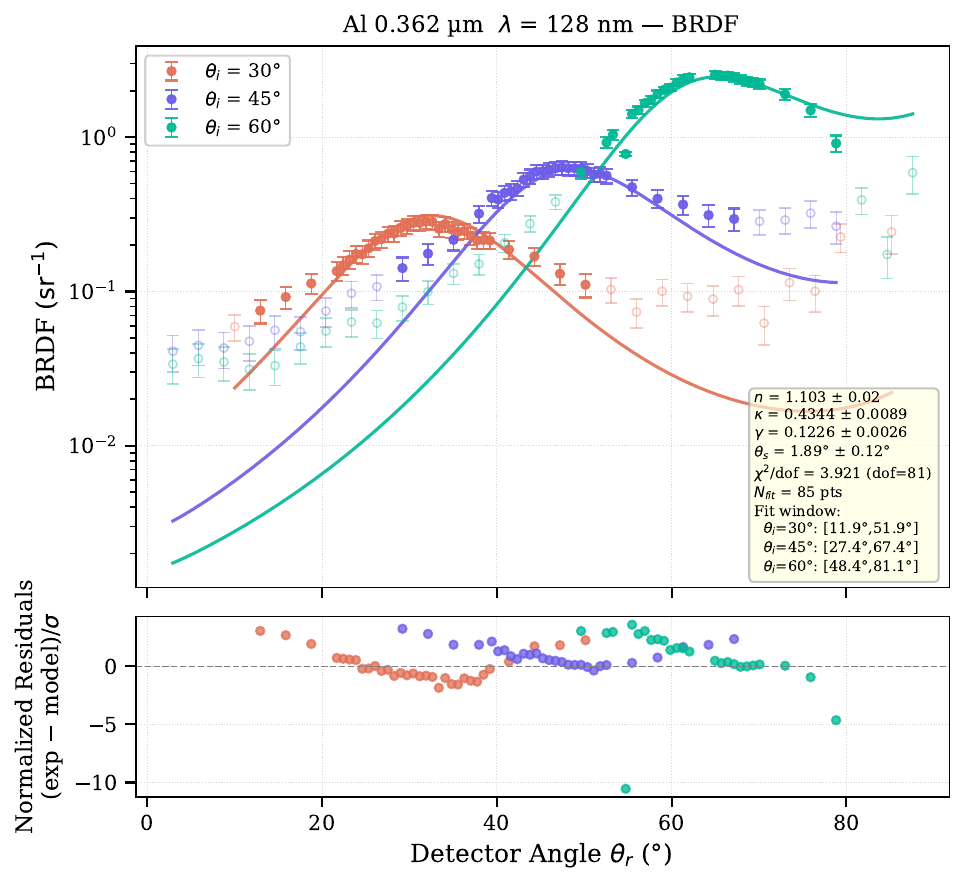}
            \caption{\qty{128}{\nm}}
            \label{fig:brdf_fit_al_128}
        \end{subfigure}\hfill
        \begin{subfigure}[b]{0.32\linewidth}
            \centering
            \includegraphics[width=\linewidth]{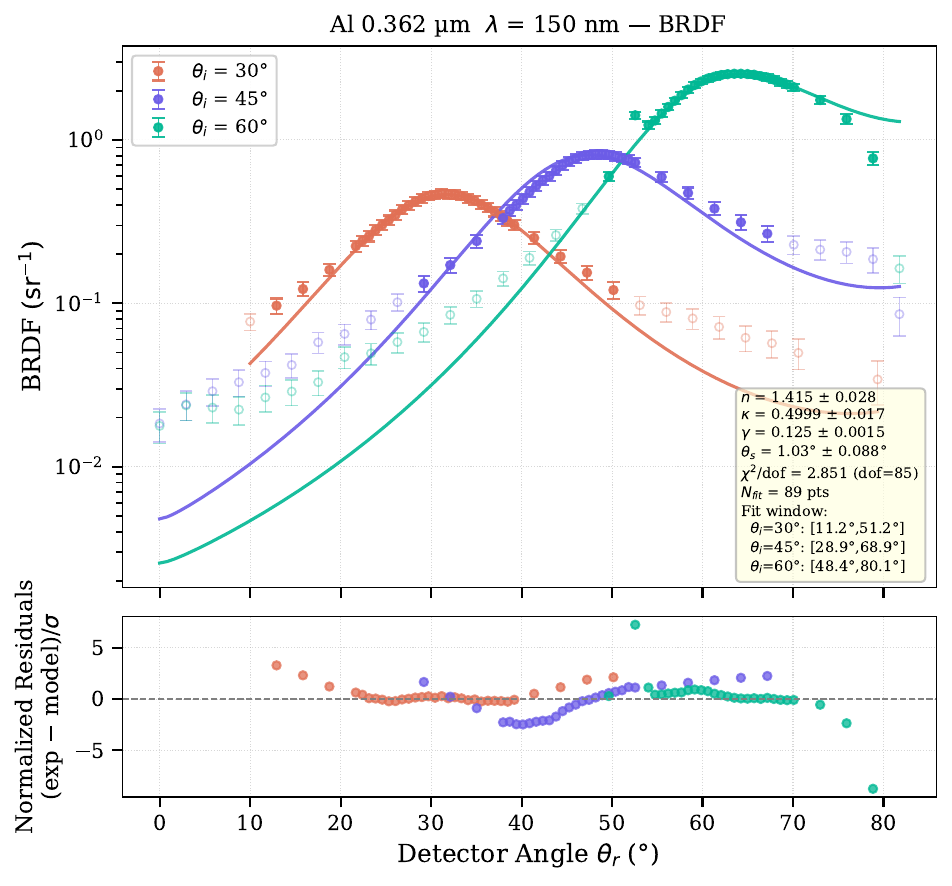}
            \caption{\qty{150}{\nm}}
            \label{fig:brdf_fit_al_150}
        \end{subfigure}\hfill
        \begin{subfigure}[b]{0.32\linewidth}
            \centering
            \includegraphics[width=\linewidth]{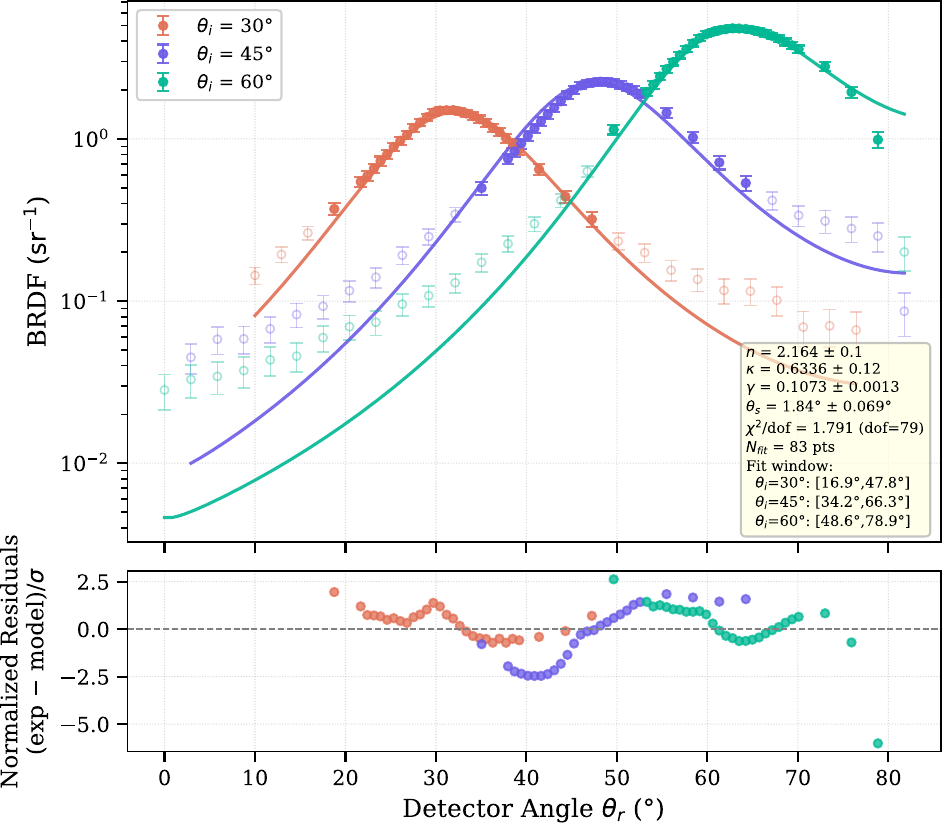}
            \caption{\qty{200}{\nm}}
            \label{fig:brdf_fit_al_200}
        \end{subfigure}

        \vspace{0.3cm}

        \begin{subfigure}[b]{0.32\linewidth}
            \centering
            \includegraphics[width=\linewidth]{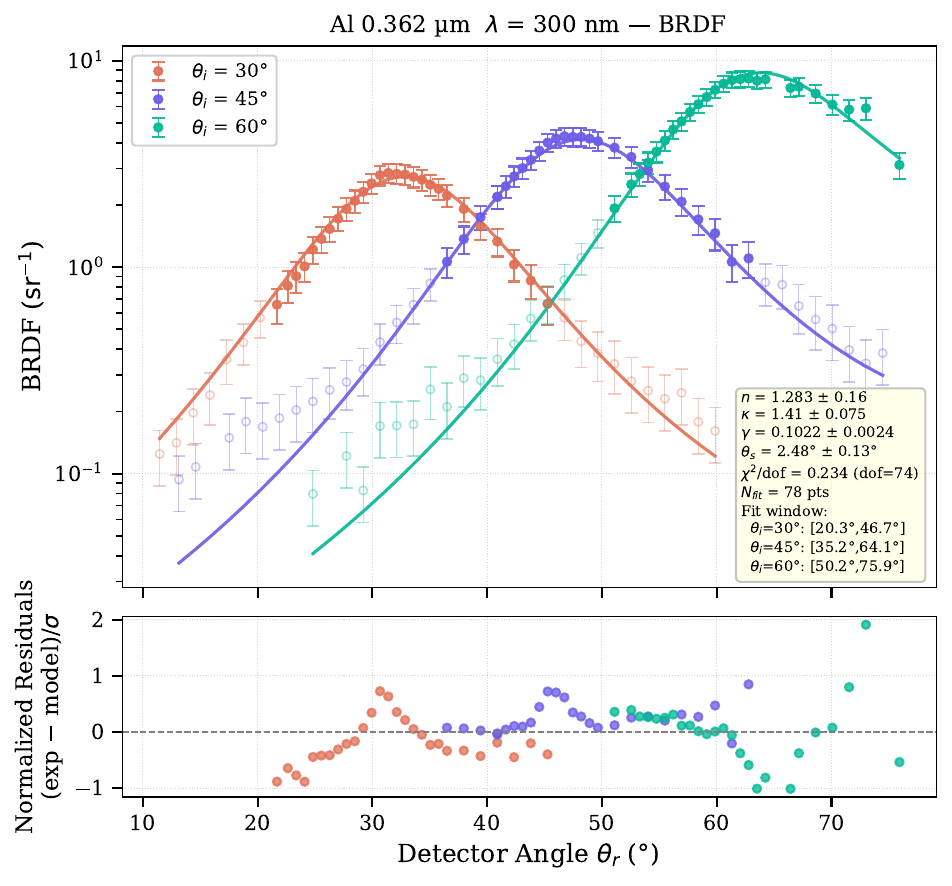}
            \caption{\qty{300}{\nm}}
            \label{fig:brdf_fit_al_300}
        \end{subfigure}%
        \hspace{0.02\linewidth}%
        \begin{subfigure}[b]{0.32\linewidth}
            \centering
            \includegraphics[width=\linewidth]{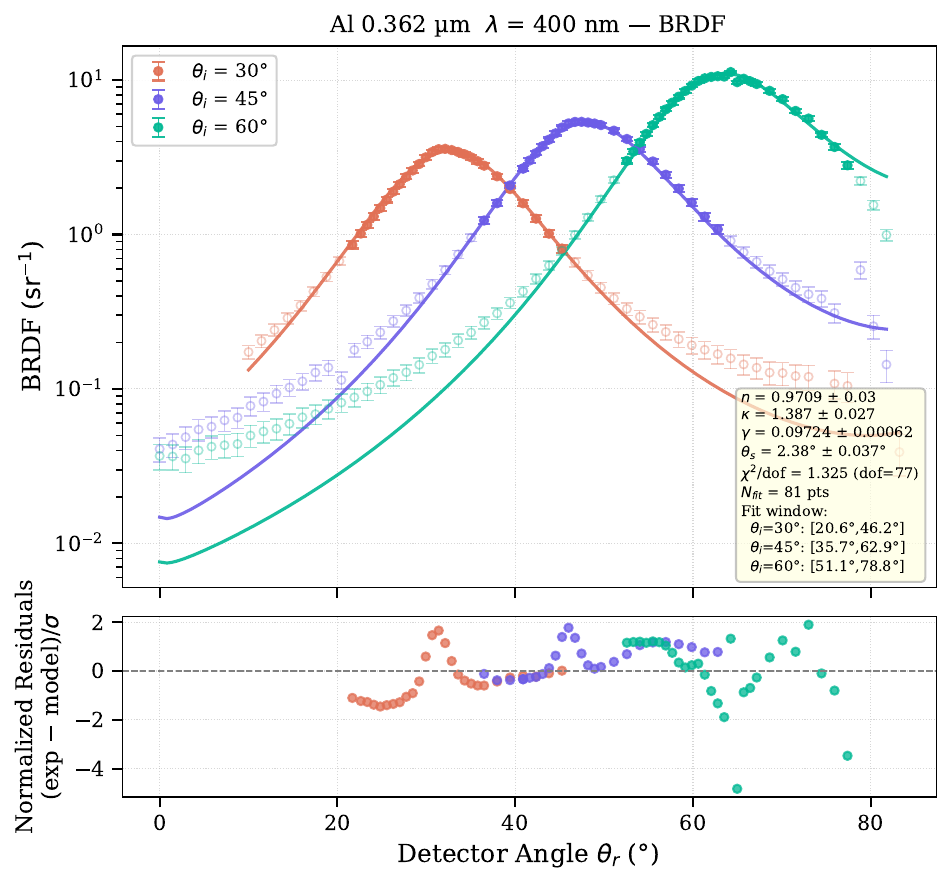}
            \caption{\qty{400}{\nm}}
            \label{fig:brdf_fit_al_400}
        \end{subfigure}
    \end{minipage}
    \caption{Fitted semilogarithmic BRDF profiles for Al \qty{0.362}{\micro\meter} at auxiliary wavelengths.}
    \label{fig:app_fits_al_semilog}
\end{figure*}

\begin{figure*}[!htbp]
    \centering
    \begin{minipage}{0.9\textwidth}
        \centering
        \begin{subfigure}[b]{0.32\linewidth}
            \centering
            \includegraphics[width=\linewidth]{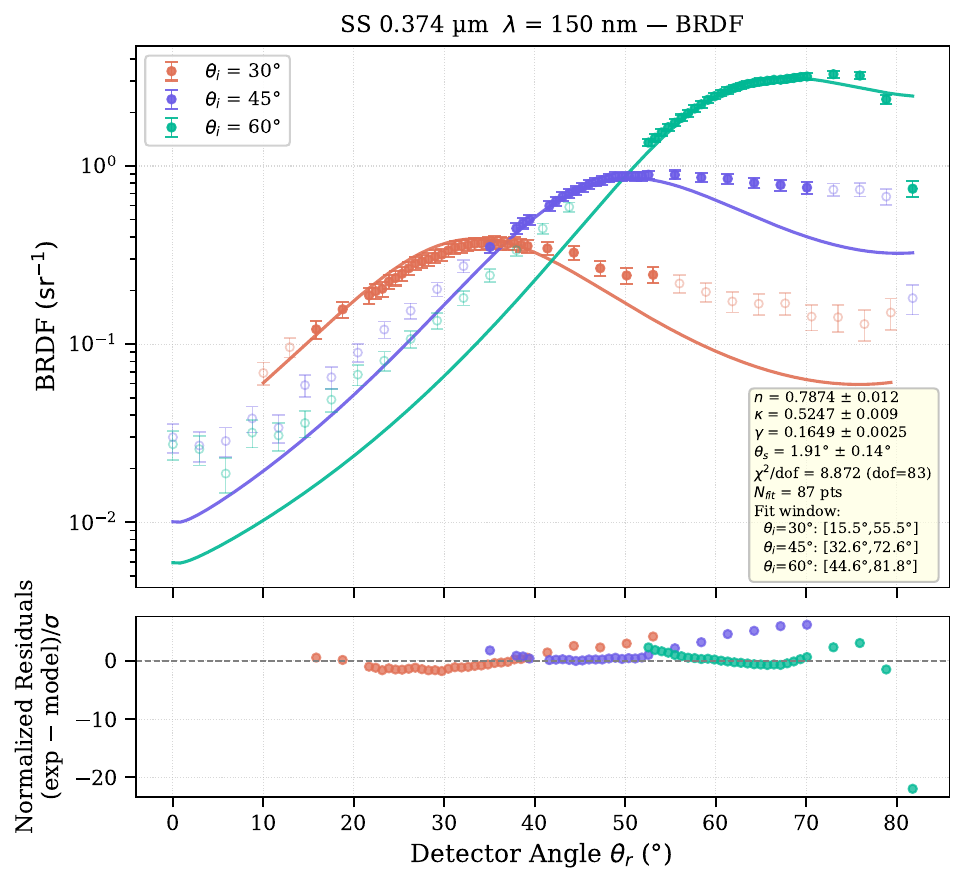}
            \caption{\qty{150}{\nm}}
            \label{fig:brdf_fit_ss_150}
        \end{subfigure}%
        \hspace{0.02\linewidth}%
        \begin{subfigure}[b]{0.32\linewidth}
            \centering
            \includegraphics[width=\linewidth]{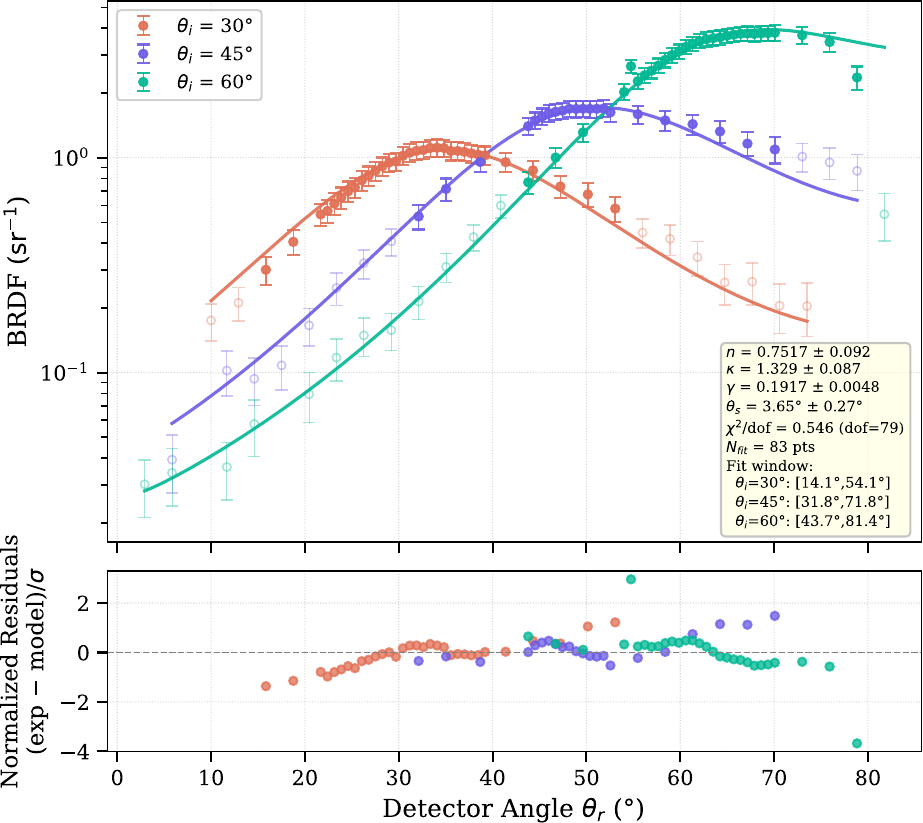}
            \caption{\qty{200}{\nm}}
            \label{fig:brdf_fit_ss_200}
        \end{subfigure}

        \vspace{0.3cm}

        \begin{subfigure}[b]{0.32\linewidth}
            \centering
            \includegraphics[width=\linewidth]{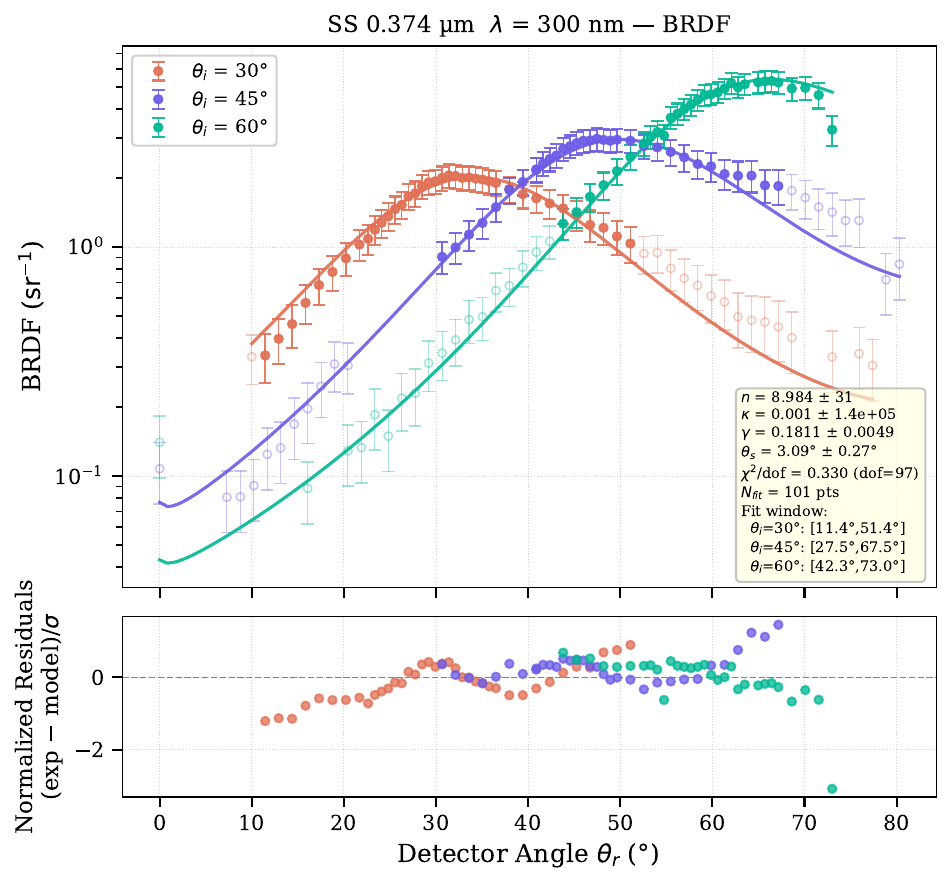}
            \caption{\qty{300}{\nm}}
            \label{fig:brdf_fit_ss_300}
        \end{subfigure}%
        \hspace{0.02\linewidth}%
        \begin{subfigure}[b]{0.32\linewidth}
            \centering
            \includegraphics[width=\linewidth]{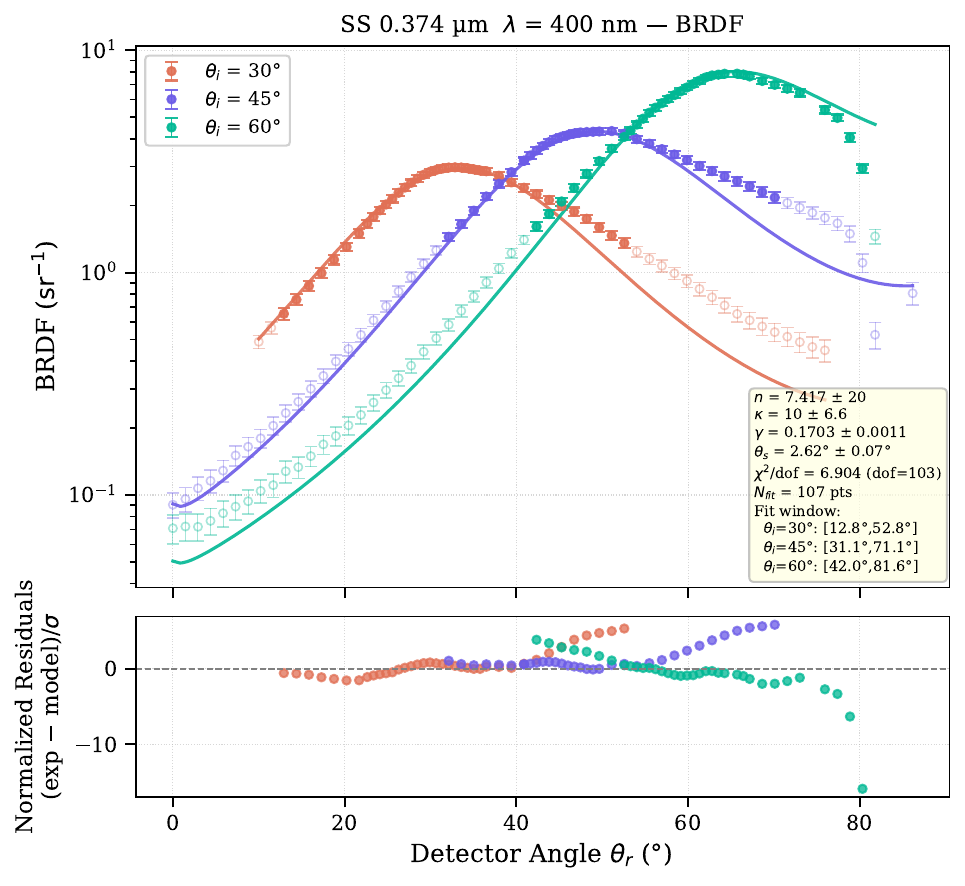}
            \caption{\qty{400}{\nm}}
            \label{fig:brdf_fit_ss_400}
        \end{subfigure}
    \end{minipage}
    \caption{Fitted semilogarithmic BRDF profiles for SS \qty{0.374}{\micro\meter} at auxiliary wavelengths.}
    \label{fig:app_fits_ss_semilog}
\end{figure*}

\Cref{fig:app_fits_al_semilog} (\Cref{fig:app_fits_ss_semilog}) shows the semilogarithmic BRDF fit results for the aluminum (stainless steel) sample at auxiliary VUV and UV-VIS wavelengths.

\subsection{Hemispherical Reflectance Curves}
\label{subsec:app_hemi_curves}

\begin{figure*}[!htbp]
    \centering
    \begin{minipage}{0.75\textwidth}
        \begin{subfigure}[b]{\textwidth}
            \centering
            \includegraphics[width=\textwidth]{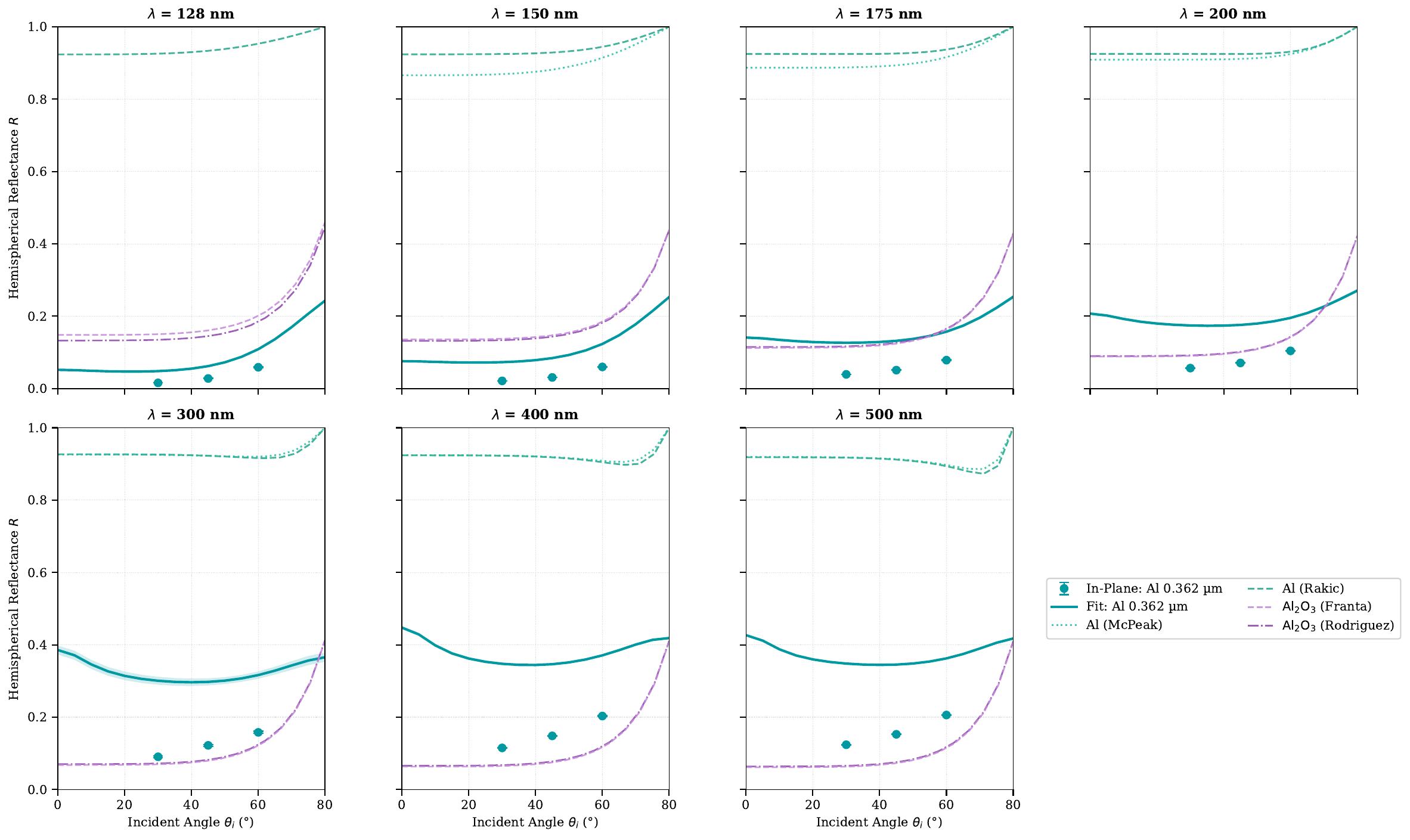}
            \caption{Al \qty{0.362}{\micro\meter}}
            \label{fig:hemi_refl_vs_aoi_al}
        \end{subfigure}
    
        \vspace{0.4cm}
    
        \begin{subfigure}[b]{\textwidth}
            \centering
            \includegraphics[width=\textwidth]{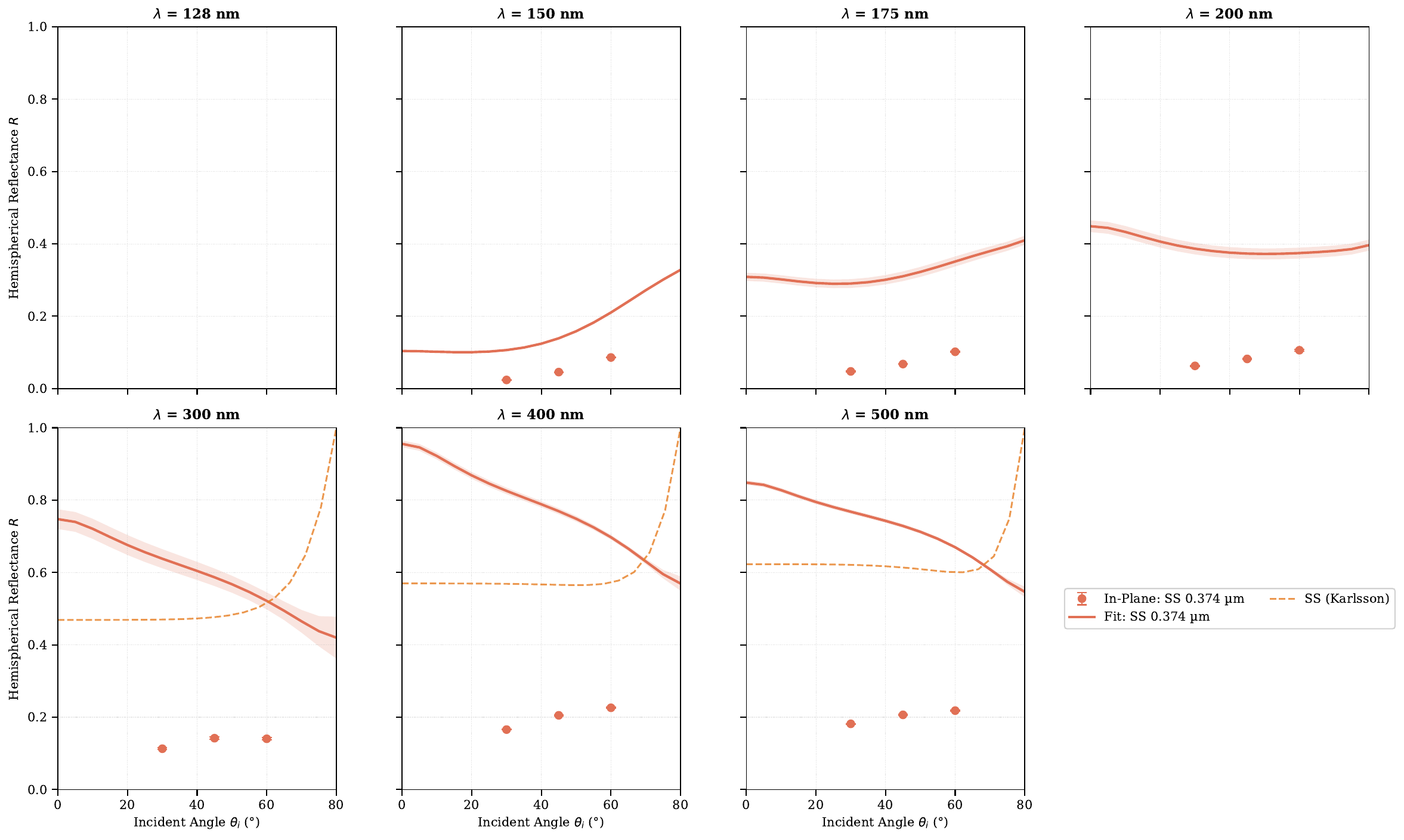}
            \caption{SS \qty{0.374}{\micro\meter}}
            \label{fig:hemi_refl_vs_aoi_ss}
        \end{subfigure}
    \end{minipage}
    \caption{Integrated hemispherical reflectance as a function of the angle of incidence for Al \qty{0.362}{\micro\meter} (a) and SS \qty{0.374}{\micro\meter} (b) across all analyzed wavelengths.}
    \label{fig:app_hemi_refl_vs_aoi}
\end{figure*}

The integrated hemispherical reflectance values as a function of the angle of incidence covering all wavelengths are depicted in \Cref{fig:app_hemi_refl_vs_aoi} for both the aluminum and stainless steel samples.

\section{Supplementary BRDF Correlation Matrices}
\label{sec:app_correlations}

\begin{figure*}[!htbp]
    \centering
    \begin{minipage}{0.9\textwidth}
        \centering
        \begin{subfigure}[b]{0.32\linewidth}
            \centering
            \includegraphics[width=\linewidth]{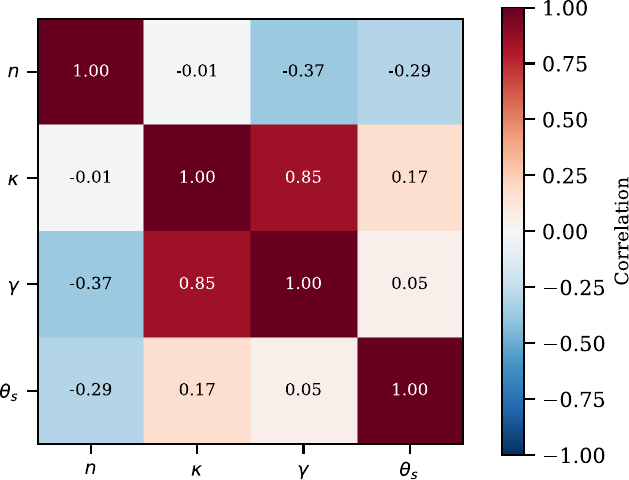}
            \caption{\qty{128}{\nm}}
            \label{fig:corr_app_al_128}
        \end{subfigure}\hfill
        \begin{subfigure}[b]{0.32\linewidth}
            \centering
            \includegraphics[width=\linewidth]{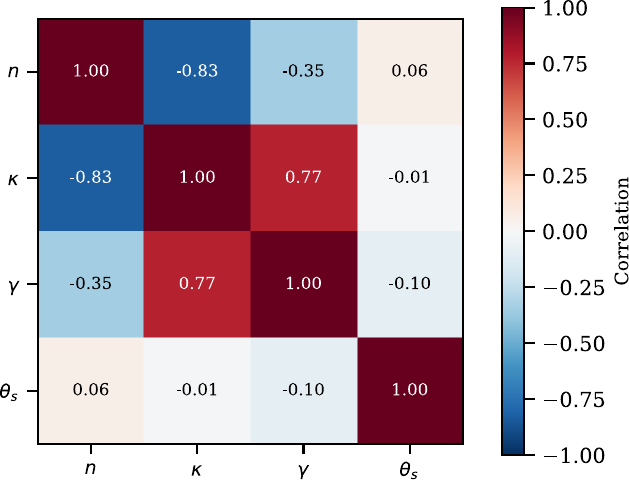}
            \caption{\qty{150}{\nm}}
            \label{fig:corr_app_al_150}
        \end{subfigure}\hfill
        \begin{subfigure}[b]{0.32\linewidth}
            \centering
            \includegraphics[width=\linewidth]{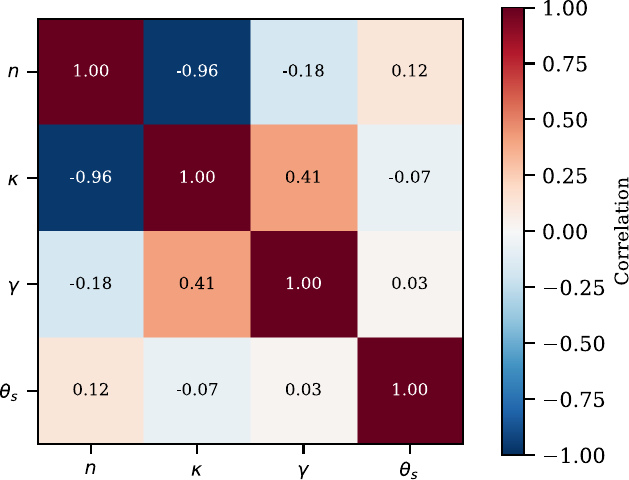}
            \caption{\qty{200}{\nm}}
            \label{fig:corr_app_al_200}
        \end{subfigure}

        \vspace{0.3cm}

        \begin{subfigure}[b]{0.32\linewidth}
            \centering
            \includegraphics[width=\linewidth]{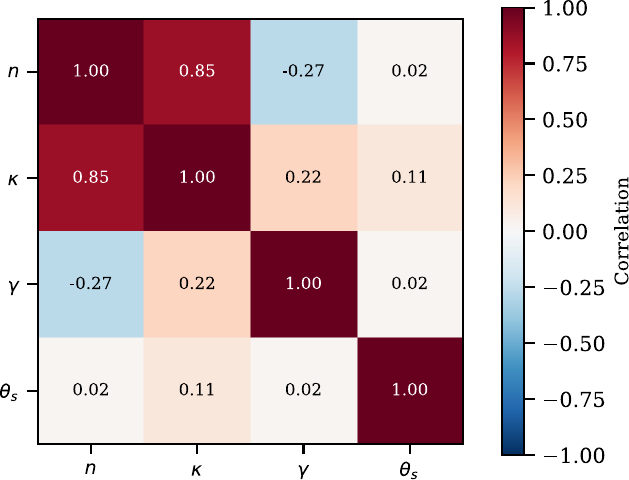}
            \caption{\qty{300}{\nm}}
            \label{fig:corr_app_al_300}
        \end{subfigure}%
        \hspace{0.02\linewidth}
        \begin{subfigure}[b]{0.32\linewidth}
            \centering
            \includegraphics[width=\linewidth]{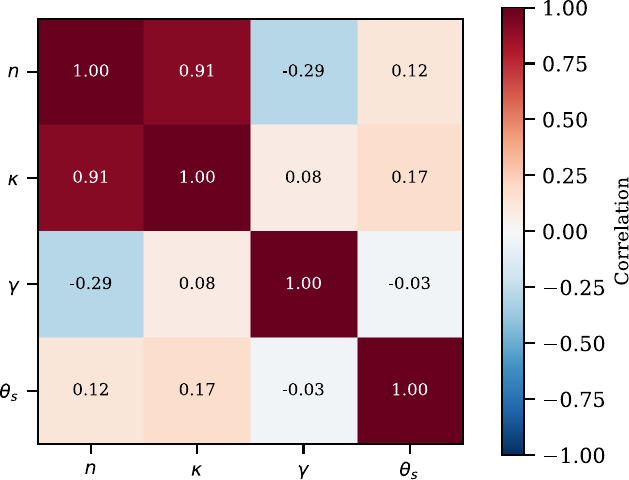}
            \caption{\qty{400}{\nm}}
            \label{fig:corr_app_al_400}
        \end{subfigure}
    \end{minipage}
    \caption{Correlation matrices between GGX model fitted parameters for Al \qty{0.362}{\micro\meter} at auxiliary wavelengths.}
    \label{fig:app_corr_al_vuv_vis}
\end{figure*}

\begin{figure*}[!htbp]
    \centering
    \begin{minipage}{0.9\textwidth}
        \centering
        \begin{subfigure}[b]{0.32\linewidth}
            \centering
            \includegraphics[width=\linewidth]{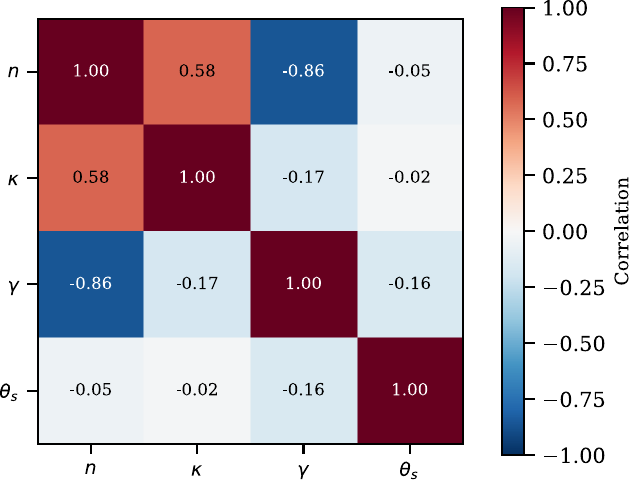}
            \caption{\qty{150}{\nm}}
            \label{fig:corr_app_ss_150}
        \end{subfigure}%
        \hspace{0.02\linewidth}%
        \begin{subfigure}[b]{0.32\linewidth}
            \centering
            \includegraphics[width=\linewidth]{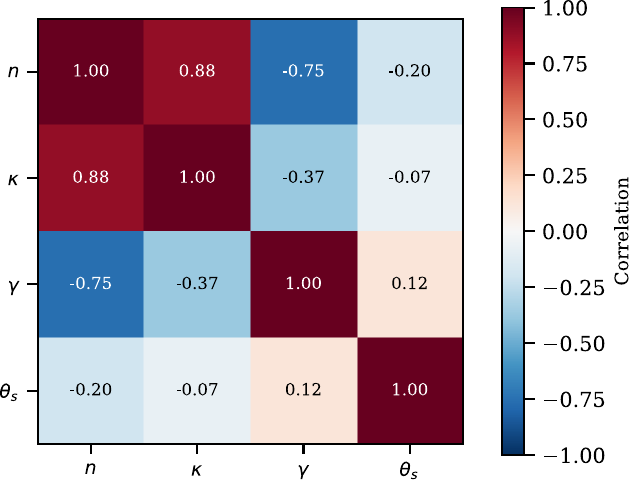}
            \caption{\qty{200}{\nm}}
            \label{fig:corr_app_ss_200}
        \end{subfigure}

        \vspace{0.3cm}

        \begin{subfigure}[b]{0.32\linewidth}
            \centering
            \includegraphics[width=\linewidth]{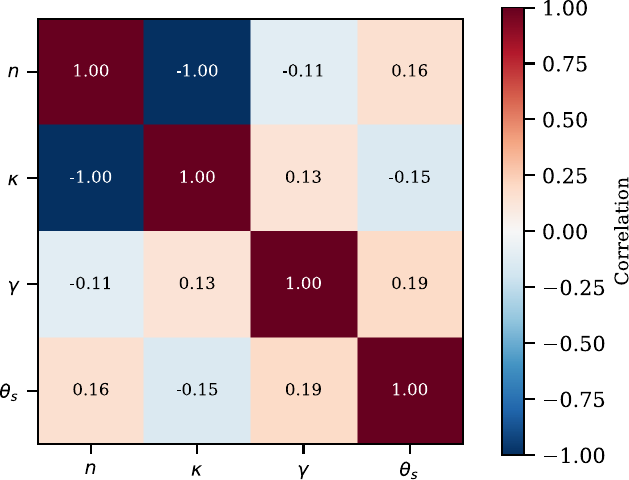}
            \caption{\qty{300}{\nm}}
            \label{fig:corr_app_ss_300}
        \end{subfigure}%
        \hspace{0.02\linewidth}%
        \begin{subfigure}[b]{0.32\linewidth}
            \centering
            \includegraphics[width=\linewidth]{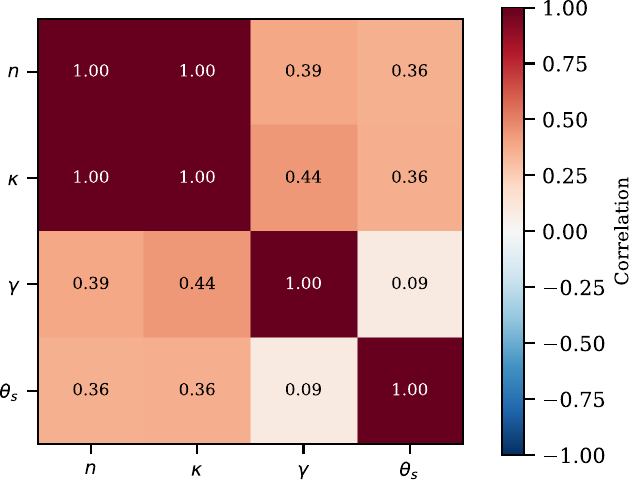}
            \caption{\qty{400}{\nm}}
            \label{fig:corr_app_ss_400}
        \end{subfigure}
    \end{minipage}
    \caption{Correlation matrices between GGX model fitted parameters for SS \qty{0.374}{\micro\meter} at auxiliary wavelengths.}
    \label{fig:app_corr_ss_vuv_vis}
\end{figure*}

This appendix presents the correlation matrices between the fitted parameters of the GGX BRDF model ($n$, $\kappa$, $\gamma$, $\theta_s$) for the remaining analyzed wavelengths (\qtylist[list-units = single]{128;150;200;300;400}{\nm}). The matrices are presented for both the aluminum (AW-1050 (A), \qty{0.362}{\micro\meter} average roughness, \Cref{fig:app_corr_al_vuv_vis}) and stainless steel (SS 304, \qty{0.374}{\micro\meter} average roughness, \Cref{fig:app_corr_ss_vuv_vis}) samples. The stainless steel measurement at \qty{128}{\nm} is omitted due to substantial gaseous argon purity drift.

\bibliographystyle{elsarticle-num}

\bibliography{references}

\end{document}